\documentstyle[11pt,a4]{article}
\font\tendl=msym10  scaled \magstep1
\font\sevendl=msym7 scaled \magstep1
\font\fivedl=msym5 scaled \magstep1
\font\tengl=eufm10  scaled \magstep1
\font\sevengl=eufm7 scaled \magstep1
\font\fivegl=eufm5 scaled \magstep1
\newfam\dlfam \def\dl{\fam\dlfam\tendl}
\textfont\dlfam=\tendl \scriptfont\dlfam=\sevendl
\scriptscriptfont\dlfam=\fivedl
\newfam\glfam 
\textfont\glfam=\tengl \scriptfont\glfam=\sevengl
\scriptscriptfont\glfam=\fivegl

\def\ifundefined#1{\expandafter\ifx\csname#1\endcsname\relax}
\makeatletter \ifundefined{new@mathgroup} {} \else
    \new@mathgroup\sffam
    \new@internalmathalphabet\mathsf\sffam{cmss}{m}{n}
    \def\psf{\fontfamily\sfdefault \fontseries\default@series
        \fontshape\default@shape\selectfont\mathsf}
\fi

\def\citen#1{\if@filesw \immediate\write \@auxout {\string\citation{#1}}\fi%
\@tempcntb\m@ne \let\@h@ld\relax \def\@citea{}%
\@for \@citeb:=#1\do {\@ifundefined {b@\@citeb}%
    {\@h@ld\@citea\@tempcntb\m@ne{\bf ?}%
    \@warning {Citation `\@citeb ' on page \thepage \space undefined}}%
    {\@tempcnta\@tempcntb \advance\@tempcnta\@ne
    \setbox\z@\hbox\bgroup\ifcat0\csname b@\@citeb \endcsname \relax
    \egroup \@tempcntb\number\csname b@\@citeb \endcsname \relax
    \else \egroup \@tempcntb\m@ne \fi \ifnum\@tempcnta=\@tempcntb
    \ifx\@h@ld\relax \edef \@h@ld{\@citea\csname b@\@citeb\endcsname}%
    \else \edef\@h@ld{\hbox{--}\penalty\@highpenalty
    \csname b@\@citeb\endcsname}\fi
    \else \@h@ld\@citea\csname b@\@citeb \endcsname \let\@h@ld\relax \fi}%
\def\@citea{,\penalty\@highpenalty\hskip.13em plus.13em minus.13em}}\@h@ld}
\def\@citex[#1]#2{\@cite{\citen{#2}}{#1}}%
\def\@cite#1#2{\leavevmode\unskip\ifnum\lastpenalty=\z@\penalty\@highpenalty\fi%
  \ [{\multiply\@highpenalty 3 #1%
  \if@tempswa,\penalty\@highpenalty\ #2\fi}]}   %
\makeatother % end of CITE.STY

\newcounter{listnr}
\def\ab            {\op\ue}
\def\aeins         {\oplus A_{1}}
\def\abar          {\mbox{$\bar\alpha$}}
\def\aeins         {\oplus A_{1}}
\newcommand{\al}[1]{\mbox{$\alpha^{(#1)}$}}
\newcommand{\andauthoretc}[5]{\\[.2 mm]{}\centerline{and}{}\\[2 mm]
                   \centerline{\sc #1}\\[2 mm] \centerline{#2}\\[.5 mm]
                   \centerline{#3}\\[.5 mm] \centerline{#4}\\[.5 mm]
                   \centerline{#5}}
\long\def\authoretc#1#2#3#4#5{\centerline{\sc #1}\\[2 mm]{}%
                   \centerline{#2}\\[.5 mm] \centerline{#3}\\[.5 mm]
                   \centerline{#4}\\[.5 mm] \centerline{#5}}
\newcommand{\arraycent}[2] {$ \left\{\begin{array}{c} #1 \end{array}
                   \right.$ \hspace{1 mm}} % {#2 mm}}

\def\ax            {\mbox{$\alpha^{(\xx)}$}}
\def\be            {\begin{equation}}
\def\bgg           {Bern\-stein\hy Gel\-fand\hy Gel\-fand}
\newcommand{\cc}[1]{\mbox{$c=#1$}}
\def\cft           {conformal field theory}
\def\cfts          {conformal field theories}
\def\Chi           {{\cal X}}
\def\clearpg       {\clearpage}
\let\cli=\centerline

\def\csa           {Cartan subalgebra}
\newcommand{\csfull}[5]{${\cal C}[ (#1)_k\oplus {\rm so}(#2)_1 \,/\,
                   (#3)_{#4} \oplus {\rm u(1)}_{#5}]$}
\newcommand{\csfulL}[5]{${\cal C}[(#1)_k\oplus {\rm so}(#2)_1 \,/\,
                   (#3)_{#4} \oplus (#3)_{#4} \oplus (#3)_{#4} \oplus
                   {\rm u(1)}_{#5}]$}
\newcommand{\csfulld}[7]{${\cal C}[(#1)_k\oplus {\rm so}(#2)_1 \,/\,
                   (#3)_{#4} \oplus(#5)_{#6} \oplus {\rm u(1)}_{#7}]$}
\newcommand{\csfullD}[7]{${\cal C}[(#1)_k\oplus {\rm so}(#2)_1 \,/\,
                   (#3)_{#4} \oplus(#5)_{#6} \oplus(#5)_{#6}
                   \oplus {\rm u(1)}_{#7}]$}
\def\csp           {\,/\,}
\def\cwb           {Cartan\hy Weyl basis}
\def\cy            {Calabi\hy Yau}
\def\cym           {Calabi\hy Yau manifold}
\long\def\coset#1#2#3{\mbox{${\cal C}[#1/ #2]_{#3}^{}$}}
\long\def\cosetk#1#2{\mbox{${\cal C}[#1/ #2]_k^{}$}}
\def\ct            {coset theory}
\def\cts           {coset theories}
\let\de=\ell
\def\delg          {\Delta^g}
\def\delgm         {\Delta_-^g}
\def\delgp         {\Delta_+^g}
\def\delgpm        {\Delta_\pm^g}

\def\delhp         {\Delta_+^h}
\def\delhpm        {\Delta_\pm^h}
\def\delwm         {\Delta_-^{[w]}}
\def\delwp         {\Delta_+^{[w]}}
\def\delwpm        {\Delta_\pm^{[w]}}
\def\dfe           {\mbox{$D$5{\tiny1}}}
\def\dfz           {\mbox{$D$5{\tiny2}}}
\long\def\del#1    \enddel{}

\def\dh            {\mbox{$\Delta_h$}}
\def\dhlind#1      {\\[-#1 mm]&&&\\ \hline\hline  {}&&&\\[-#1 mm] }
\def\dhline#1      {\\[-#1 mm]&\\ \hline\hline  {}&\\[-#1 mm] }
\def\dhlinf#1      {\\[-#1 mm]&&&&&\\ \hline\hline  {}&&&&&\\[-#1 mm] }
\def\dhlinv#1      {\\[-#1 mm]&&&&\\ \hline\hline  {}&&&&\\[-#1 mm] }
\def\dhlinz#1      {\\[-#1 mm]&&\\ \hline\hline  {}&&\\[-#1 mm] }
\def\dm            {\mbox{$\Delta_-$}}
\def\dpl           {\mbox{$\Delta_+$}}
\long\def\drac#1#2{{\displaystyle\frac{#1}{#2}}}
\def\dyd           {Dynkin diagram}
\def\ee            {\end{equation}}

\let\emb=\hookrightarrow
\def\emt           {energy-momentum tensor}
\def\epop          {extended Poin\-car\'e polynomial}
\newcommand{\erf}[1]{(\ref{#1})}
\newcommand{\fline}[1]{\\[#1 mm]\noindent ------------------\\[1 mm]}
\def\futnote#1     {\footnote{~#1}\ }
\def\gid           {\mbox{${\cal G}_{\rm id}$}}
\def\Gl            {\mbox{$G$2{\tiny1}}}
\def\gm            {\mbox{g}}
\def\gs            {\mbox{${\cal G}_{\rm s}$}}
\def\Gs            {\mbox{$G$2{\tiny2}}}
\def\gv            {\mbox{$g_{}^\vee$}}
\def\h             {\mbox{$K$}}
\def\hing          {\mbox{$h\emb g$}}
\def\hiv           {\mbox{$h_i^{\vee}$}}
\def\hq            {\mbox{$\cal Q$}}
\def\hsc           {hermitian symmetric coset}
\def\hy            {$\mbox{-\hspace{-.66 mm}-}$}
\def\ic            {\mbox{$I_{\rm c}$}}
\def\ii            {{\rm i}}
\def\jc            {\mbox{$J_{\rm c}$}}
\def\jf            {J.\ Fuchs}
\def\js            {\mbox{$J_{\rm s}$}}
\def\jv            {\mbox{$J_{\rm v}$}}
\def\kma           {Kac\hy Moo\-dy algebra}
\def\ks            {Ka\-za\-ma\hy Su\-zu\-ki}
\newcommand{\la}[1]{\mbox{$\Lambda_{(#1)}$}}
\def\labar         {\mbox{$\tilde\Lambda^{{}^{\scriptstyle\bar\alpha}}_{}$}}
\def\labarw        {\mbox{$\tilde\Lambda_{[w]}^{{}^{\scriptstyle\bar\alpha}}$}}
\def\laful         {{\tilde\lambda}}
\def\lafull        {\mbox{$\tilde\lambda$}}
\def\lafulp        {{{\tilde\lambda}_{}^{+_{}}}}
\def\lagi          {Landau\hy Ginz\-burg}
\def\lie           {Lie algebra}
\def\Lie           {Lie group}
\def\lx            {\mbox{$\Lambda_{(\xx)}$}}
\def\mo            {{M_s}}

\def\nabar         {\mbox{$:\!\!\Psi^{\bar\alpha}_{}
                                \Psi^{-\bar\alpha}_{}\!\!:$}}
\def\ng            {\mbox{$\delta N$}}

\def\ngen          {\mbox{$N_{27}^{}$}}
\def\nagen         {\mbox{$N_{\overline{27}}$}}
\def\nn            {$N=2$\ }
\def\nnn           {$N=2$}
\def\NS            {Neveu\hy Schwarz\ }
\def\nxe           {\setcounter{listnr}{1} \thelistnr}
\def\nxt           {\addtocounter{listnr}{1} \thelistnr}
\def\onehalf       {\mbox{$\frac12$}}
\let\op=\oplus
\def\opop          {ordinary Poin\-car\'e polynomial}
\newcommand{\pcs}[2]{$(A, 1, #1, #2)$}
\newcommand{\pf}[4]{\mbox{$\Phi^{#1,{\rm #2}}_{#3,#4}$}}
\newcommand{\pF}[3]{\mbox{$\Phi^{#1,{\rm #2}}_{#3}$}}
\def\pop           {Poin\-car\'e polynomial}
\def\ptx           {\mbox{${\cal P}(t,x)$}}
\def\Q             {\mbox{$\tilde v_{\circ}$}}
\def\qq            {\mbox{$\xi_\circ$}}
\def\qm            {\mbox{$Q_{\rm m}$}}
\def\qsuco         {\mbox{$q_{\rm suco}$}}
\def\qx            {\mbox{$Q_{\xx}$}}
\def\R             {Ramond }
\long\def\rank#1   {\mbox{rank}\,#1}

\def\Rep           {Representation}
\def\rhog          {\mbox{$\rho_g$}}
\def\rhoh          {\mbox{$\rho_h^{}$}}
\def\rgs           {Ramond ground state}
\newcommand{\sect}[1] {\section{#1}\setcounter{equation}{0}}
\def\sign          {\mbox{sign\,}}
\def\sod           {\mbox{so(2$d$)}}
\def\sode          {\mbox{so(2$d$)$_1$}}
\def\Sord          {\mbox{$|\gid|$}}
\def\spc           {\hspace{4.6 em}}
\def\spd           {\hspace{2.4 em}}
\def\sumabar       {\mbox{${\displaystyle\sum_{\bar\alpha\in\Delta_+}}$}}
\def\sumabaR       {\mbox{$\sum_{\bar\alpha\in\Delta_+}$}}

\newcommand{\titleetc}[6]{{\tt #1}\mbox{}\\[-21mm]\rightline{{\sf\prepnr}}
     {}\\[.7 mm] \rightline{{\sf hep-th/9304133}} {}\\[.7 mm]%
     \rightline{{\sf #2}} {}\\[13 mm]%
     \cli{\bf\Large {#3}}\\[2.1 mm] \cli{\bf\Large {#4}}\\[10 mm]{}%
     {#5} \\[7 mm]{}%
     \begin{quote}{\bf Abstract.}\ \,{#6}\end{quote} \newpage}%
\def\twodim        {two-di\-men\-si\-o\-nal}
\def\ue            {\mbox{u(1)}}
\def\v             {\mbox{$v_{\circ}$}}
\def\vol           {{\rm vol}}
\def\wg            {\mbox{$W_g$}}
\def\wgh           {\mbox{$W_g/W_h$}}
\newcommand{\Wgh}[2]{\mbox{$W(#1)/W(#2)$}}
\def\wh            {\mbox{$W_h$}}
\newcommand{\wmax}[1]{\mbox{$w_{\rm max}^{{#1}}$}}
\def\wmx           {\mbox{$w_{\rm max}^{}$}}
\def\wzwt          {WZW theory}
\def\wzwts         {WZW theories}
\def\xx            {i_\circ}
\def\zet           {{\dl Z}}

\def\zetpluso      {\mbox{${\zet}_{\geq 0}$}}

\newcommand{\mpcis}[6] {\multiput(#1,#2)(#3,#4){#5}{\circle*{#6}}}
\newcommand{\mplin}[8] {\multiput(#1,#2)(#3,#4){#5}{\line(#6,#7){#8}}}

\newcommand{\mpsss}[6] {\multiput(#1,#2)(#3,#4){#5}{{\scriptsize #6}}}

\newcommand{\pulin}[5] {\put(#1,#2){\line(#3,#4){#5}}}

\newcommand{\putss}[3] {\put(#1,#2){{\scriptsize #3}}}

 \newcommand{\wb}       {\,\linebreak[0]}
   \newcommand{\wB}       {$\,$\wb}
   \newcommand{\J}[1]     {{{#1}}\vyp}
   \newcommand{\JJ}[1]    {{{#1}}\vyp}
   \newcommand{\Bi}[1]    {\bibitem{#1}}
   \newcommand{\Prep}[2]  {preprint {#1}}

   \newcommand{\BOOK}[4]  {{\em #1\/} ({#2}, {#3} {#4})}
   \newcommand{\vyp}[4]   {\ {#1} ({#2}) {#3}}
   
   \newcommand{\amst} {Amer.\wb Math.\wb Soc.\wb Transl.}
   \newcommand{\anma} {Ann.\wb Math.}
   \newcommand{\anop} {Ann.\wb Phys.}
   \newcommand{\comp} {Com\-mun.\wb Math.\wb Phys.}
   \newcommand{\ijmp} {Int.\wb J.\wb Mod.\wb Phys.\ A}
   \newcommand{\kniz}[2] {in: {\em The Physics and Mathematics of Strings,
              Memorial Volume for V.G.\ Knizhnik}, {L.\ Brink et al., eds.}
              (\WS, \Si 1990), {{#1}}{{#2}} }
   \newcommand{\lemp} {Lett.\wb Math.\wb Phys.}
   \newcommand{\mams} {Memoirs\wB Amer.\wb Math.\wb Soc.}
   \newcommand{\mpla} {Mod.\wb Phys.\wb Lett.\ A}
   
   \newcommand{\npbp} {Nucl.\wb Phys.\ B (Proc.\wb Suppl.)}
   \newcommand{\nupb} {Nucl.\wb Phys.\ B}
   \newcommand{\phlb} {Phys.\wb Lett.\ B}
   \newcommand{\rmap} {Rev.\wb Math.\wb Phys.}
   
   \newcommand{\AP}     {{Academic Press}}

   \newcommand{\CUP}    {{Cambridge University Press}}

   \newcommand{\SV}     {{Sprin\-ger Verlag}}
   
   \newcommand{\WS}     {{World Scientific}}
   \newcommand{\Be}     {{Berlin}}
   \newcommand{\Ca}     {{Cambridge}}
   \newcommand{\NY}     {{New York}}

   \newcommand{\Si}     {{Singapore}}
\newcommand{\labl}[1]{\label{#1}\ee}

\def\prepnr{{HD-THEP-93-9}} 
\begin{document} \titleetc{}
 {April 1993} {NON-HERMITIAN SYMMETRIC $N=2$ COSET MODELS,}
   {POINCAR\'E POLYNOMIALS, AND STRING COMPACTIFICATION}
 {\authoretc{\ J\"urgen Fuchs\ $^1$}{NIKHEF-H}{Kruislaan 409}
   {NL -- 1098 SJ~~Amsterdam}{}
 \andauthoretc{\ Christoph Schweigert\ $^2$}{Institut f\"ur
   Theoretische Physik}{Philosophenweg 16} {D -- 6900~~Heidelberg}{}}
 {The field identification problem, including fixed point resolution,
is solved for the non-hermitian symmetric \nn superconformal \cts.
Thereby these models are finally identified as well-defined modular
invariant \cfts. As an application, the theories are used as subtheories
in \nn tensor products with \cc9, which in turn are taken as the inner
sector of heterotic superstring compactifications. All string theories
of this type are classified, and the chiral ring as well as the number
of massless generations and anti-generations are computed with the help
of the \epop. Several equivalences between a priori different non-hermitian
\cts\ show up; in particular there is a level-rank duality for an infinite
series of \cts\ based on $C$ type \lie s. Further, some general results
for generic \nn \cts\ are proven: a simple formula for the number of
identification currents is found, and it is shown that the set of \rgs s
of any \nn coset model is invariant under charge conjugation.
\fline {9} $^1$~~Heisenberg fellow\\[.4 mm]
$^2$~~supported by Studienstiftung des deutschen Volkes}

\sect{Introduction}
While the conditions necessary for the
consistency of a superstring theory seem to be too weak to pinpoint a
`theory of everything', string theory remains an interesting approach
to unify the fundamental interactions including
gravity. Furthermore, the study of strings has given new and deep insight
in various topics in mathematics and physics so that there are good
reasons, beyond possible direct application to phenomenology, to have
a closer look at the structures arising in string theory.

A class of \twodim\ field theories for which this point of view is
particularly justified are the \nn superconformal theories which
are needed for the inner sector of (heterotic) string theories.
The enlarged (\nnn) world sheet supersymmetry for the right-moving
part of the theory is in this case dictated \cite{sen.,bdfm}
by the requirement of space-time supersymmetry, a property imposed for
phenomenological reasons, such as to `solve' the gauge hierarchy problem.
In the present paper, we consider theories for which \nn supersymmetry
is present in the left-moving part as well;
just like in the generic case, these \nn theories are interesting
in their own right, as they are singled out by the presence of
new structures such as the ring of chiral primary fields and
the connection with \cym s \cite{levw}. Furthermore, there exist
deep relations between \nn super\cfts\ and \cfts\ in general, including
the interpretation of the fusion ring of any rational \cft\ as
a deformation of the chiral ring of some \nn theory
\cite{gepn9,intr2,lewa2,cres}.

There exist several approaches to construct the inner sector of a heterotic
string theory: non-linear sigma models with a \cy\ manifold as their
target space \cite{chsw}, the description in terms
of \lagi\ potentials \cite{grvw,mart2}, and exactly solvable models
(these approaches are closely interrelated, but the question
to which extent they are equivalent has not yet been
resolved completely). By exactly solvable we mean that all correlation
functions can (at least in principle) be calculated exactly. Among
the solvable super\cfts\ there are free field constructions employing
the Coulomb gas approach \cite{dofa3},
and theories constructed by algebraic methods.
In the algebraic approach the {\em coset construction\/} \cite{goko2}
plays a prominent r\^ole, for it allows to obtain many superconformal
theories within the framework of affine \kma s.

In \cite{kasu2} Kazama and Suzuki considered coset models $\cal C$
of the form
  \be \cosetk{g\op\sod}h\,.    \labl{kasuform}
Here $g$ stands for a semi-simple Lie algebra, and
$h$ is a reductive subalgebra of $g$; the integer $d$ is defined
as $2d = \dim g - \dim h$, while the integer
$k$ denotes the level of the affinization $g^{(1)}_{}$ of $g$.
As shown in \cite{kasu2}, the symmetry algebra of such coset models
a complete list of all \nn \cts\ of the form (\ref{kasuform})
was obtained. Indeed the following conditions are necessary and
sufficient for a \ct\ of the form \erf{kasuform}
to have \nn superconformal symmetry:
\begin{enumerate}
\item The embedding $h \hookrightarrow g$ has to be regular.
\item The number
  \be   n:=\onehalf\,(\rank g - \rank h)  \ee
must be an integer.
\item Denoting the simply connected compact Lie groups having $g$ and $h$
as their Lie algebras by $G$ and $H$, respectively, the coset {\em manifold\/}
   \be     \frac{G}{H \times \mbox{U}(1)^{2n}}       \ee
has to be K\"ahlerian.  \end{enumerate}

\noindent Up to now the following theories solving these constraints
have been considered in the literature: \begin{itemize}
\item Tensor products of \nn minimal models \cite{gepn3},
including models which employ non-diagonal
and non-product modular invariants \cite{scya4,fkss2,scya7,lysc3,fksv}.
\item Tensor products of the so-called projective cosets \cite{foiq},
 corresponding to \cts\  of the form
 \be \cosetk{\mbox{su}(n+1)\op\mbox{so}(2n)_1}{\mbox{su}(n)\op\ue}\,. \ee
  For these models non-diagonal modular invariants have been investigated,
  too \cite{aafn}.
\item Tensor products of arbitrary hermitian symmetric \cts\ (`HSS-cosets')
  with the diagonal modular invariant \cite{sche3}.
\end{itemize}
Note that \nn minimal models can be considered as projective cosets
with $n=1$, and
projective cosets are a subclass of the hermitian symmetric cosets.

{}From the classification \cite{kasu,schW} of \nn superconformal coset models
in the \ks\ framework it is well known that there exist even more
models that possess \nn superconformal symmetry; the hermitian symmetric
\cts\ constitute only
a subclass. In this paper we shall consider the general case.
The paper is organized as follows. First, we recall in section 2
the classification of \nn\
superconformal coset models obtained in the \ks\ framework. As a by-product
we prove a simple characterization of hermitian symmetric spaces which
differs from the one given in the standard literature.
Based on the general classification, we then
provide a complete list of all non-\hsc s that
can be used in tensor products with conformal central charge $c=9$.

We proceed by specifying the \cfts\ defining the cosets of our interest.
This is necessary because a `Lie-algebraic coset' $\cal C$
as it stands in \erf{kasuform}\
is in itself far from defining a consistent modular invariant \cft.
We emphasize that although the theories described in this paper have been
introduced as formal cosets \erf{kasuform} already five years ago,
they have previously {\em not\/} been
shown to describe consistent \cfts. (By consistency of a conformal
field theory we understand among other requirements that the characters
of the theory carry a (projective) unitary representation of the modular
group. Note that up to now
it is even unknown whether a conformal field theory can be
associated with every coset, and if so, whether this theory is unique.)
To define the theories, we first determine the precise form of the
affinization of the subalgebras involved. In particular we identify,
in section 3.1,
the level of the \ue-subalgebra that is present in each of the models.

Moreover,
as is well known \cite{gepn8,scya5}, in order to obtain a modular
invariant partition function, `fields' in the coset theory have to be
`identified'.
Problems arise when the length of the identification orbits is not
constant; orbits of non-maximal length have to be `resolved'
\cite{scya5,scya6}, which in general is a rather delicate issue.
It is common to assume \cite{scya5} that the field identifications
can be reduced to purely group-theoretical selection rules
(see however the counter example found in \cite{dujo}).
In section 3.2 of our paper we determine these identification
rules.  Furthermore we derive a formula, valid for any \nn \ct\
of the form \erf{kasuform}, for the order of the abelian group that
is generated by the identification currents; this provides
a convenient check for the completeness of the identification rules.
The resolution of fixed points is dealt with in section 3.3.

In section 4 we apply the results of section 3 to string
compactification. We first calculate in section 4.1 the ring of
chiral primary fields \cite{levw} of the models.
To be able to read off the \pop\ and (part of) the spectra of
massless particles of the models,
we derive in section 4.2 a formula giving the full
superconformal \ue-charge of any \rgs\ in terms of the length of
an associated element of the Weyl group. (This length is conveniently
calculated by means of Hasse diagrams; the diagrams corresponding to our
models are described in appendix A.)
After presenting the results for the \pop s, we also show, in
section 4.3, that the set of Ramond ground states of any \nn \ct\
is symmetric under charge conjugation.

In section \ref{44} we present the complete list
of all tensor products of \cts\ that involve at least one non-hermitian
symmetric \ct\ and have central charge $c=9$, providing
thus consistent vacua for heterotic string compactification to
four space-time dimensions \cite{gepn5}.
When combined with the list of tensor products involving only minimal models
\cite{fkss2} and with the corresponding list for hermitian symmetric spaces
\cite{foiq}, this completes
the list of all tensor products of \nn \cts\ that can be
obtained from cosets of the type (\ref{kasuform}). Note that
the set of all string vacua is much bigger than the set of all
tensor products of \cts, as in general by choosing different modular
invariants of the $g$- and $h$-\,\wzwts\ one gets
different string vacua. However, to obtain this set is, at
present, beyond reach, as a complete classification of modular invariants
is still lacking for WZW theories based on simple \lie s other than $A_1$.

The spectra of these compactifications are computed with the
help of the \epop\ that, as described in section \ref{44},
can be deduced from the \opop\ and the action of the so-called
spinor current. Sometimes the results obtained for the \epop s of
a priori different \cts\ are identical, which suggests
that the corresponding \cfts\ are closely related and maybe even
identical. In section 4.5 we present an infinite series of
theories for which this phenomenon occurs and describe a map that
provides a one to one correspondence between
the primary fields of the respective theories.

Finally, in section 5 we conclude with a brief summary and an outlook
on possible further work.

\sect{Classification}
In \cite{kasu2} a supersymmetric extension of the coset
construction \cite{goko2} was used to obtain a large
class of superconformal coset models. By bosonizing
the fermions of the super \wzwts\ involved in the construction of
these models, one arrives at a level one \sode\ \wzwt. As a
consequence, the models can be written as
  \be  \cosetk{g \op\sode}h \,  .     \labl{2.1}

In the sequel we will adopt the notation of \cite{kasu} and denote indices
referring to generators of the algebra $g$ by capital letters $A, B, \ldots$\,,
indices referring to the subalgebra $h$ by $a, b, \ldots$\,, and indices
referring to the set $g\setminus h$,
and hence also to \sod, by $\bar{a}, \bar{b}, \ldots\;$.
Thus in particular the currents generating $g$ are denoted by $\hat J^A$,
and the \sod\ algebra is generated by $\dim g/h$ fermions $j^{\bar{a}}$.
Denoting the structure constants of $g$ by $f^{AB}_{\ \ C}$, the currents
  \be \tilde J^a = \hat J^a - \frac\ii k\, f^a_{\ \bar b \bar c}\,
  j^{\bar b} j^{\bar c}        \labl{embed}
then specify the embedding of $h$ in $g\op\sod$.
 \futnote{Unless stated otherwise, we use the summation convention,
i.e.\ equal upper and lower indices should be contracted.}
{}From the embedding (\ref{embed}) we can read off the levels of the
simple subalgebras $h_i$ of
  \be  h=\hat{h}\op\ue^m_{} = \bigoplus_i h_i \op\ue^m_{} .  \labl{hhat}
Namely,
  \be k(h_{i}) = I_i\, (k +\gv) -\hiv,  \labl a
where \gv\ and \hiv\ denote the dual Coxeter numbers of $g$ and $h_i$,
respectively, and where $I_i$ is the Dynkin index
of the embedding $h_i \hookrightarrow g$, i.e.\ the relative length
squared of the highest roots $\theta_g$ of $g$ and $\theta_i$ of $h_i$,
   \be  I_i := \frac{(\theta_g, \theta_g)}{(\theta_i,\theta_i)} . \ee
(Here and below we often
follow the habit of referring to an untwisted affine \kma\
$f^{(1)}$ by its horizontal subalgebra $f$, and to the Heisenberg
algebra $\hat{\rm u}(1)$ by its horizontal subalgebra \ue, whenever no
confusion can arise. In particular, the reductive subalgebra
$h$ will sometimes stand for
its affinization $\bigoplus_i h_i^{(1)} \op\hat{\rm u}(1)^m$.
Also, we use the short hand notation $f_k$ if $f^{(1)}$ is at
level $k$.) With \erf a, the conformal central charge of the \ct\ becomes
  \be  c= \frac32 \,(\dim g - \dim h) - \frac{(\theta_g,\theta_g)\, \gv\dim g
  - \sum_i(\theta_i,\theta_i)\,\hiv\dim h_i } {(\theta_g,\theta_g)(k+\gv)}.
  \labl{26}

The symmetry algebra of the models \erf{2.1} always contains the $N=1$
supersymmetry algebra. To find a second supercurrent $G^{2}$,
one starts with the most general ansatz expressing a spin $3/2$
current of the coset theory \erf{kasuform} in terms of the
currents $\hat J^A$ and the fermions $j^{\bar a}$ \cite{kasu},
  \be  G^2(z) =\frac2k\, (\h_{\bar{a}\bar{b}} :\!j^{\bar{a}}\hat{J}
  ^{\bar{b}}\!: - \frac\ii{3k} S_{\bar{a} \bar{b} \bar{c}} :\!j^{\bar{a}}
  j^{\bar{b}} j^{\bar{c}}\!: ) \,.       \ee
Here the colons denote normal ordering, and $S$ is a totally antisymmetrical
tensor. This ansatz mimics the structure of the first supercurrent $G^{1}$
for which $\h_{\bar{a} \bar{b}}$ and $ S_{\bar{a} \bar{b} \bar{c}}$ are given
by the Killing form $\kappa_{\bar{a} \bar{b}}$ and by the structure
constants $f_{\bar{a} \bar{b} \bar{c}}$, respectively.

The calculation of the relevant operator products that involve
$G^2(z)$ shows that the following set of equations for \h\ and $S$
is necessary and sufficient for enlarged supersymmetry:
  \begin{eqnarray}
 \h_{\bar{a}\bar{b}} & = & - \h_{\bar{b}\bar{a}}, \qquad
 \h_{\bar{a}\bar{b}} \h_{\bar{b} \bar{c}} =
 - \delta_{\bar{a} \bar{c}}  \label{ss1} ,    \\
 \h^{\bar{a} \bar{b}} f_{\bar{b} \bar{c} e} & = &
 f^{\bar{a} \bar{b}}_{\ \;e} \h_{\bar{b} \bar{c}}, \label{ss2} \\
 f_{\bar{a} \bar{b} \bar{c}} & = &
 \h_{\bar{a} \bar{p}} \h_{\bar{b} \bar{q}}
 f^{\bar{p} \bar{q}}_{\ \;\bar{c}}  \,+\,
   \mbox{cyclic permutations in } \bar{a},\bar{b}
   \mbox{ and } \bar{c},\label{ss3}  \\
 S_{\bar{a} \bar{b} \bar{c}} & = & \h_{\bar{a} \bar{p}}
 \h_{\bar{b} \bar{q}} \h_{\bar{c} \bar{r}}
 f^{\bar{p} \bar{q} \bar{r}}.  \label{ss4} \end{eqnarray}
The condition (\ref{ss1}) means that \h\ is a complex structure on
$G/H$, which is $h$-invariant
by (\ref{ss2}). (\ref{ss3}) is a consistency condition, while
(\ref{ss4}) can be used to eliminate $S$ from the problem.

This set of equations can also be understood in more geometrical
terms. Namely, let $t$ denote the orthogonal complement of $h$
with respect to the Killing form $\kappa$ of $g$ (this is well
defined since, $g$ being semi-simple, $\kappa$ is non-degenerate).
Then the model \cosetk{g\op\sode}h is \nn supersymmetric if and
only if there exists a direct sum decomposition of vector spaces,
 \be t =  t_+ \op t_- ,   \labl{zerl}
which obeys the conditions that $\dim{t_+} = \dim{t_-}$, that
$t_+ $ and $t_-$ separately form closed Lie algebras, and
that the restriction of the Killing form to $t_+$ and to $t_-$ vanishes,
  \be \kappa_{\mid t_{\pm}}  \equiv 0 . \labl{solv}

This geometric characterization is in fact rather easy to prove \cite{kasu}.
Suppose first that the theory
\cosetk{g\op\sode}h is \nn supersymmetric. Define  $ t_{\pm} $ to
be the eigenspaces corresponding to the eigenvalues $ \pm \ii $ of the
complex structure \h.  Then the relations $ t = t_{+} \oplus t_{-}$
and $ \dim {t_{+}} = \dim{t_{-}} $ are immediate.
Using (\ref{ss1}) to (\ref{ss4}), it is also easy to show that
  \be  [ t_{\pm}^{\bar{a}} , t_{\pm}^{\bar{b}} ] = \frac12\, (\ii f^{\bar{a}
  \bar{b}}_{\ \;\bar{c}} \pm S^{\bar{a}\bar{b}}_{\ \;\bar{c}} )\;
  t_{\pm}^{\bar{c}}, \ee
where $t_\pm^{\bar{a}} $ denotes the component of $ t^{\bar{a}} $ in
$t_\pm$. Thus the elements of $t_\pm$ close under the Lie bracket.
Finally, for arbitrary $r_\pm,\, s_\pm \in t_\pm$
the antisymmetry (\ref{ss1}) of \h\ implies
 $\kappa (r_\pm , s_\pm) = \mp\ii\, \kappa (\h r_\pm, s_\pm) = \pm\ii\,
  \kappa (r_\pm ,\h s_\pm) =- \kappa (r_\pm , s_\pm) = 0 $,
so that \erf{solv} holds.
Conversely, given a decomposition like (\ref{zerl}),
define \h\ by requiring   $ t_{\pm} $ to be the eigenspaces of \h\
corresponding
to the eigenvalues $ \pm \ii$, assuring that  the second equation of
(\ref{ss1}) is fulfilled. Then (\ref{ss2}), (\ref{ss3}) can be shown to
follow from the fact that
$ t_{\pm} $ are subalgebras, while (\ref{solv}) implies the first part of
(\ref{ss1}). Namely, for arbitrary $ r,s \in t $ one has
$ r = r_{+} + r_{-} $ and
$ s = s_{+} + s_{-} $ with  $ r_{\pm}, s_{\pm} \in t_{\pm} $, and
therefore
$  \kappa (\h r , s) = \kappa (\ii r_{+}-\ii r_{-} , s_{+}+s_{-}) =
\ii \kappa (r_{+} , s_{-}) - \ii \kappa (r_{-} , s_{+}) = -\kappa
(r_{+}+r_{-} , \ii s_{+}-\ii s_{-}) = - \kappa ( r , \h s ) $.

Our task is now to classify embeddings satisfying \erf{ss1} to
\erf{ss4}, or, equivalently, \erf{zerl} and \erf{solv}.
As the following remarks show, we can assume that $g$ and $h$ are of
equal rank. In \cite{kasu} a sequential method has been
introduced which allows us to reduce \nn \cts\ with
rank\,$h < $  rank\,$g$ to the equal rank case.
(It is worthwhile mentioning that the validity of this sequential
algorithm has been proven in \cite{kasu} only as far as the symmetry
algebras of the models are concerned. As for
the field contents, the general belief is that for a chain of
embeddings $f\hookrightarrow h\hookrightarrow g$ the \ct\ \coset gf{}
carries the structure of the tensor product of the theories \coset gh{} and
\coset hf{}, albeit a non-product modular invariant must be used.
This is easy to see if no
field identification is necessary, and should also hold in the case when
the identification currents do not have fixed points.)
To apply the sequential method, one needs an intermediate subalgebra satisfying
  \be h \subseteq h \op\ue^{\mbox{\scriptsize rank\,$g-$rank\,$h$}}
  \subseteq g \labl{chain}
(direct sum of Lie algebras).
Such an intermediate algebra exists only \cite{schW}
for the so-called regular subalgebras. A regular subalgebra \hing\
is by definition (see e.g.\ \cite{HUmp}) a subalgebra for which
every generator associated to a root of the subalgebra $h$ is also
associated to a root of the overlying algebra $g$; all other subalgebras
are called special.
In \cite{schW} it was  shown that the cosets derived from special
subalgebras  never have enlarged supersymmetry; correspondingly
we can restrict ourselves in the sequel to regular subalgebras,
and hence the sequential algorithm is applicable. Regular subalgebras
have been classified by Dynkin \cite{dynk}; their Dynkin diagram
must be a subdiagram of the extended Dynkin diagram
of the overlying algebra (the extended Dynkin diagram of a simple
\lie\ $g$ coincides with the \dyd\ of its affinization $g^{(1)}$).

In short, we can restrict our attention to regular embeddings
satisfying rank\,$g=$\,rank\,$h$.
We now turn to the classification of such embeddings
generating \nn superconformal \cts. {}From the \nn conditions (\ref{ss1})
to (\ref{ss4}), one easily deduces that
  \be f^{cde} \,\h_{\bar{a} \bar{b}}^{} \, f^{\bar{a} \bar{b}}_{\ \;e}
  = 0 \labl{hproj}
for all $c, d$.
We will denote by \dpl, \dm, and \dh\ the sets of roots of $t_{+},$ $t_{-}$,
and $h$, respectively, and define
  \be \Q := \sumabar \abar.   \labl Q
Writing (\ref{hproj}) in a \cwb\ and comparing prefactors, we find
  \be  (\Q,\gamma) = 0 \quad \mbox{ iff }\; \gamma \in \dh.  \labl{hpro}
This relation implies that
  \be [ \sumabar \abar_i H^i , T^a] = 0
  \quad \mbox{ for all } \quad T^{a} \in h,  \labl{aht}
where by $H^i$ we denote the generators of the \csa, i.e.\ that
$h$ contains a \ue\ ideal with generator $\sumabaR \abar_i H^i$.
Thus the embedding \hing\ is such that the Dynkin diagram
of $h$ is obtained from the extended Dynkin diagram of $g$ by
removing at least two nodes.
One can also show \cite{kasu} that
  \be (\Q,\bar\beta) \geq (\bar{\beta},\bar\beta) > 0     \labl{Absch}
for all $\bar{\beta} \in \dpl$.

We claim that the subalgebras yielding \nn superconformal cosets
are precisely {\em diagram subalgebras}, i.e.\ subalgebras whose \dyd\
is contained in the {\em non-extended\/} Dynkin diagram of $g$.
Moreover, if the \dyd\ of $h$ is obtained from that of $g$ by
removing more than one node, then the sequential method alluded to
above can be applied to reduce the theory to a tensor product;
hence we can assume that only a single node is deleted.
We will denote by $\xx$ the label of this distinguished node of
the Dynkin diagram of $g$; thus, for example,
\ax\ is the corresponding simple $g$-root that is not a
root of $h$. Note that the notation \Q\ introduced in \erf Q
was chosen with foresight; for instance,
denoting the fundamental $g$-weights by \la i,
the relation (\ref{hpro}) can be rephrased as
  \be  \Q \propto \lx  \ee
(the constant of proportionality,
obtainable with the help of the strange formula, reads
  \be  \frac{ (\theta_g, \theta_g)\, \gv \dim g - \sum_i
  (\theta_i, \theta_i)\, h_i^\vee \dim h_i }
  { 12 \sum_j G_{\xx j}} \, , \labl{strange}
where $G_{ij}=(\la i,\la j)$ denotes the metric on the weight space of $g$,
i.e.\ the inverse of the symmetrized Cartan matrix).

To prove the above claim,
we have to show that the highest root $\theta_g$ of $g$ is not
a root of $h$. If $\theta_g$ were a root of $h$, then according to
\erf{hpro} it would satisfy $(\Q,\theta_g)=0$. But this is not
allowed, as can be seen with the help of
the decomposition of $\theta_g$ in terms of the simple $g$-roots \al i,
  \be \theta_g = \sum_{i=1}^{{\rm rank}\, g} a_i\, \al i  .   \labl{howu}
Namely, the coefficients $a_i$ on the right hand side of \erf{howu}, known as
the Coxeter labels of $g$, are positive integers, and hence
the inequality (\ref{Absch})  implies
$(\Q,\theta_g)=\sum_{i} a_{i}\, (\Q, \al i) \geq \sumabaR
a_i (\abar^{(i)}, \abar^{(i)}) > 0.$
Thus $\theta_g$ is not a root of $h$, so that $h$ is a diagram
subalgebra of $g$.

The converse is seen as follows. Given a diagram subalgebra $h$ of $g$,
assign the root \abar\ of $t$ to belong to
$ \Delta_+$ and $\Delta_-$, respectively, iff it is a positive
respectively a negative root of $g$.
Since we assumed that $g$ and $h$ have equal rank,
this prescription yields a decomposition of $t$ of the form
(\ref{zerl}). It is now straightforward to check that the vector
spaces generated by the elements corresponding to
$\Delta_{\pm}$ satisfy the geometrical formulation of the  \nn
conditions. Namely, nilpotency (\ref{solv}) is immediate from the
well-known properties of the Killing form in a \cwb; the dimensions
of $t_+$ and $t_{-}$coincide because positive and negative roots of
$g\setminus h$ come in pairs;
and the assertion that $t_{\pm}$ close under the Lie bracket can
be verified by using the
fact that $(\Q,\bar{\alpha}) > 0$ iff $\abar\in\dpl$.

Clearly, the \nn conditions (\ref{ss1}) to (\ref{ss4}) are
particularly simple if the structure constants
$f_{\bar{a} \bar{b} \bar{c}}$ vanish. As we will see shortly, the
corresponding coset manifold is then a hermitian symmetric space.
In this case we automatically have rank\,$h$ = rank\,$g$. Moreover
using the Jacobi identity together with the relation
 \be  2  f^{\bar{a} \bar{c} d}   f_{\bar{b} \bar{c} d}
  = f^{\bar{a} C D} f_{\bar{b} C D} =
  \gv\delta^{\bar{a}}_{\,\bar{b}} ,  \ee
it is easy to show that
  \be f_{\bar{c} \bar{d}e} \,\h_{\bar{a} \bar{b}}\, f^{\bar{a} \bar{b} e} =
 \gv \,\h_{\bar{c} \bar{d}}.   \labl{gproj}
Similarly as with (\ref{hpro}), another useful relation is obtained
by writing (\ref{gproj}) in a \cwb; comparing prefactors one finds
  \be (\Q,\bar\gamma)\equiv\sumabar (\abar,\bar{\gamma}) = \gv \quad
  \mbox{ iff } \; \bar{\gamma} \in \dpl.  \labl{gpro}

With these results, we are in a position to classify all subalgebras
yielding hermitian
symmetric spaces. Let us first sketch the way these spaces are usually
described in the mathematical literature (see e.g.\ \cite{HElg}).
Given $f_{\bar{a} \bar{b } \bar{c}} = 0$, it is possible to define
an involutive automorphism $\sigma$ of the \lie\ $g$ such that the
subalgebra left invariant by $\sigma$ is equal to $h$, namely $\sigma (
T^{a}) := T^{a}$, $\sigma(T^{\bar{a}}) := - T^{\bar{a}}$.
\lie s admitting such an automorphism are called orthogonal involutive
\lie s and have been classified by Cartan;
a complete list can be found e.g.\ in  \cite[p.\ 354]{HElg}.
Because of \erf{aht}, among the orthogonal involutive \lie s one only
has to consider those whose fixed algebra contains a \ue\ ideal.
Finally, one verifies by inspection that for all such \lie s
the \nn conditions are fulfilled.

(The nomenclature used above arises from the following geometrical
setting.  The fact that $g$ and $h$ form an orthogonal involutive
\lie\ can be shown to be equivalent to the property that the homogeneous
space $G/H$, with
$G$ and $H$ the compact simply connected Lie groups corresponding
to $g$ and $h$, is a {\em Riemannian globally symmetric space}. These spaces
are defined as follows. For a Riemannian manifold, a neighbourhood
of any point $p$ of the manifold can be described by mapping a sphere in the
tangent space at $p$ on the neighbourhood; via this map the
reflection about the origin of the tangent space (the pre-image of $p$)
induces a mapping $\tau$ of this neighbourhood.
If $\tau$ is an isometry, the manifold is called a locally
symmetric space; if in addition $\tau$ can be extended to a global
isometry, the manifold is called a globally symmetric space.  It
can be shown that all globally symmetric spaces are
homogeneous spaces, i.e.\ isomorphic to the quotient of a simply
connected Lie group by a closed subgroup.
In this geometrical context the condition \erf{aht} means
that $G/H$ carries in addition an almost
complex structure $J$ which is {\em hermitian}, i.e.\ the metric $\gm$
satisfies $\gm(JX, JY) = \gm(X, Y)$ for all elements $X$, $Y$ of the
tangent space. It can be shown that for homogeneous
spaces this automatically implies that $J$ is K\"ahlerian, i.e.\
covariantly constant.
In the general case where $f_{\bar a\bar b\bar c}$ is non-vanishing
(which is the situation in which we are interested in the present
paper), the homogeneous space $G/H$ is no longer
a Riemannian globally symmetric space, but as was shown in \cite{kasu},
it is nonetheless still a K\"ahlerian space iff the \nn conditions
are fulfilled. We remark that for our
purposes these geometric characterizations are of little use. In fact,
one of the main achievements of the theory of homogeneous spaces
was precisely to recast the problems in purely Lie algebraic terms, which
finally provided a powerful handle on the geometric objects.)

\begin{table}[t]
\caption{Hermitian symmetric \cts\ (HSS) and their Virasoro charges}
\label{HSS} \begin{center} \begin{tabular}{|l|r|l|} \hline
   &&\\[-3 mm]
\multicolumn{1}{|c|}{\cosetk gh } & \multicolumn{1}{c|}{$c$} &
\multicolumn{1}{c|}{name}\dhlinz{3.7}
\cosetk{ A_{m+n-1} }{ A_{m-1} \op A_{n-1} \ab}& $3kmn / (k+m+n)$ &
($A$, $m$, $n$, $k$)  \\
   &&\\[-3 mm]
\cosetk{ B_{n+1} } {B_n \ab} & $3k(2n+1) /(k+2n+1)$ & ($B$, $2n+1$, $k$) \\
   &&\\[-3 mm]
\cosetk{ D_{n+1} } {D_n \ab} & $6kn / (k+2n)$  & ($B$, $2n$, $k$)  \\
   &&\\[-3 mm]
\cosetk{C_n}{ A_{n-1}\ab}  & $3kn(n+1) / 2(k+n+1)$ & ($C$, $n$, $k$)  \\
   &&\\[-3 mm]
\cosetk{D_n}{ A_{n-1}\ab}  & $3kn(n-1) / 2(k+n-2)$ & ($D$, $n$, $k$)  \\
   &&\\[-3 mm]
\cosetk{E_6}{D_5\ab}    & $48k / (k+12)$   & ($E6$, $k$)      \\ &&\\[-3 mm]
\cosetk{E_7}{E_6\ab}    & $81k / (k+18)$   & ($E7$, $k$)       \\
   [.7 mm] \hline \end{tabular} \end{center} \end{table}

Alternatively, the classification of hermitian symmetric spaces
can be found by the following simple
prescription \cite{ekmy}: the hermitian symmetric spaces are obtained
by deleting a node of the \dyd\ of $g$ that corresponds to
a so-called \cite{fuge} {\em cominimal\/} fundamental weight,
i.e.\ a fundamental $g$-weight \la i\ such that $a_{i} =1$ in
the decomposition \erf{howu} of the highest $g$-root  $\theta_g$.
To prove this characterization, we proceed as follows.
Multiplying both sides of (\ref{howu}) with
\Q\ as defined in \erf Q, one obtains
  \be (\Q, \theta_g) = \sum_{i=1}^{{\rm rank}\, g} \sumabar
  a_i\, (\abar , \al i) .    \labl{holad}
Now suppose that $\theta_g$ is a root of $h$. Then according to (\ref{hpro})
one has $\sum_{\bar\alpha\in\Delta_+}
(\abar,\theta_g) = 0.$ Given the fact that the Coxeter labels
$a_{i}$ are positive, we thus learn from \erf{holad} that
$(\Q, \al i)=0 $ for all simple roots.
But then (\ref{hpro}) and (\ref{gpro}) imply that all simple roots of
$g$ are contained in  $h$, and hence $g=h$,
showing that the coset would be trivial in this case.
In conclusion, $\theta_g$ cannot be a root of $h$.
{}From \erf{gpro} we then learn that the left hand side of (\ref{holad})
equals \gv. The right hand side can take this value only in the case
when exactly one simple root of $g$ with  Coxeter label equal to 1
is not contained in $h$. Now using the classification of regular subalgebras
\cite{dynk}, it is straightforward to check that one obtains in this way
exactly the same list as before.

\begin{table}[t] \caption{Non-hermitian symmetric \cts\ relevant for
\cc9 tensor products}
\label{series} \begin{center} \begin{tabular}{|l|r|l|} \hline &&\\[-3 mm]
\multicolumn{1}{|c|}{\cosetk gh } & \multicolumn{1}{c|}{$c$} &
\multicolumn{1}{c|}{name} \dhlinz{3.7}
\cosetk{B_{n}}{A_{n-1} \ab} &
$ \drac{3}{2}\,n(n+1)-\drac{3 n^{3}}{k+2n-1}    $ & ($BA, n, k$)  \\
   &&\\[-3 mm]
\cosetk{B_{n}}{B_{n-2} \aeins \ab} &
$ 12n-15-\drac{24 (n-1)^{2}}{k+2n-1}    $ & ($BB, n, k$)  \\
   &&\\[-3 mm]
\cosetk{C_{n}}{C_{n-1} \ab} &
$ 6n-3-\drac{6 n^{2}}{k+n+1}    $ & ($CC, n, k$)  \\
   &&\\[-3 mm] \hline &&\\[-3 mm]
\cosetk{C_{3}}{A_{1} \aeins \ab} & $ 21 - \drac{75}{k+4}$ &
($C3, k $)          \\
   &&\\[-3 mm]
\cosetk{C_{4}}{A_{2} \aeins \ab} & $ 36 - \drac{162}{k+5}$ &
($C4, k $)          \\
   &&\\[-3 mm]
\cosetk{D_{4}}{A_{1} \aeins \aeins \ab} & $ 27 - \drac{150}{k+6}$ &
($D4, k $)          \\
   &&\\[-3 mm]
\cosetk{D_{5}}{A_{2} \aeins \aeins \ab} & $ 45 - \drac{324}{k+8}$ &
($\dfe, k $)          \\
   &&\\[-3 mm]
\cosetk{D_{5}}{A_{3} \aeins \ab} & $ 39 - \drac{294}{k+8}$ & ($\dfz, k $)  \\
   &&\\[-3 mm]
\cosetk{F_{4}}{C_{3} \ab} & $ 45 - \drac{384}{k+9}$ &
($F4, k $)          \\
   &&\\[-3 mm]
\cosetk{G_{2}}{A_{1}^{>} \ab} & $ 15 - \drac{50}{k+4}$ &
($\Gl, k $)          \\
   &&\\[-3 mm]
\cosetk{G_{2}}{A_{1}^{<} \ab} & $ 15 - \drac{54}{k+4}$ &
($\Gs, k $)          \\
   [2.6 mm] \hline \end{tabular} \end{center} \end{table}

In table \ref{HSS} we recall the list of all HSS models and
their Virasoro charges
(the short-hand notation displayed in column 3 is taken from \cite{foiq}).
We now return to the general case. Let us stress that
we are in a position to give a complete list of {\em all\/} \nn
coset models. However, even when grouping these
theories (of which there are infinitely many)
into a finite number of series, this list still
remains rather long, and we will not present it here
in full detail.  Rather, we list only those models
that can be used as factor theories in tensor products with conformal
central charge $c=9$ (as well as some other models which fall into
infinite series that contain models relevant for \cc9).
The interest in these models comes from superstring
theory where they can be used for the inner sector of heterotic
string vacua \cite{gepn5}, and from the possible relation with
\cy\ manifolds and with \lagi\ theories.

The result of our classification is presented in table \ref{series},
where we supply the \cts\ together with their conformal central charge
(as calculated according to \erf{26}) and with
a short-hand name that derives from the Lie algebras involved.
{}From the classification of regular subalgebras described above,
the relevant embedding \hing\ is determined uniquely by the
pair $g$, $h$ of Lie algebras
for all entries in table \ref{series} except for the two models
with $g = G_2$. In the latter cases we use the superscripts
\mbox{`\,$<$\,'} and \mbox{`\,$>$\,'} to indicate that the
$A_1$-subalgebra corresponds  to the short and long simple
root of $G_2$, respectively.

For convenience we have grouped some models in the table
in three series. {}From the
above remarks it should be clear that there is no physical
distinction between the models within these series and the other models.
The different appearance is a mere artefact of our string theory-oriented
condition on the central charges.
We also emphasize that the list in table \ref{series}
does {\em not\/} contain
all \nn coset theories with central charge $c\leq 9$. Their number is
much larger, but most of them cannot be combined with other known
\nn theories to obtain $c=9$ tensor product theories.
For instance, we have not included the model \,\cosetk{D_6}{D_4 \aeins
\ab }, which has $ c = 51 - \frac{486}{k + 10}$ . For level $k=1$
the conformal central charge is $c =\frac{75}{11}<9$, but there does
not exist any
\nn model with $c=\frac{24}{11}$ which could be tensored with this
theory to arrive at a $c=9$ \cft.

Note that the number of the models so obtained is relatively small.
This can be traced back to two simple facts. First,
if $g$ is a \lie\ of $A$ type, all subalgebras lead to \cts\
of the HSS type. Second, for any fixed \lie\ $g$,
the central charge of the \ct\ grows rather fast when one
moves the node with label $\xx$ away
from the `margin' of the Dynkin diagram of $g$ towards the
inner part (note that except for $A_r$
all cominimal fundamental weights, i.e.\
those leading to \hsc s, correspond to marginal nodes).

\sect{Specification of the \cts}
As already emphasized, the `Lie-algebraic coset' as it stands
in (\ref{kasuform}) is in itself
far from defining a consistent modular invariant conformal field theory.
In this section we will provide a detailed specification of the \cft.

In fact, the first step to do so was already taken in the previous section
when we computed the levels (\ref{a}) of the semi-simple part of the
subalgebra $h$, i.e.\ of the simple ideals in the decomposition
\erf{hhat}, which in the case of our interest reads
  \be  h=\hat{h}\op\ue = \bigoplus_i h_i \op\ue .  \ee
But the abelian ideal of $h$ must be specified as well.

\subsection{The \ue\ subalgebra}
The \cft\ corresponding to a \ue\ algebra has Virasoro charge $c=1$.
As all $c = 1$ \cfts\ have been classified \cite{dvvv,bumt}
and their field contents is known, it is sufficient to have
a look at the conformal dimensions occuring in the \cft\ we are after,
which, as we shall show now, in turn are fixed by the embedding.

The direction of the \ue\ in root space is given by \Q. {}From the
embedding \erf{embed} we read off the precise form of the \ue-generator
\hq; it is proportional to
  \be \tilde{\cal Q}(z):= (\Q, H(z)) + \sumabar (\Q, \abar) \,\nabar(z)\,.
  \labl{U1}
Here \nabar\ denotes the fermion number operator for the
complex fermion that is associated to the root $\bar{\alpha}$;
it takes integer
values in the \NS sector and half-integer values in the Ramond sector; $H$
stands for the \csa\ currents of $g$.

By replacing \Q\ in (\ref{U1}) by an appropriate multiple \v\ of \Q,
all eigenvalues of \hq\ can be taken to be
integers. We will assume that we have chosen the smallest multiple
fulfilling this requirement (otherwise we would be forced
later on to introduce additional identification currents that have a
non-trivial component only in the \ue\ part), and write
   \be     \Q\equiv\sumabar \abar = \qq\,  \v ;   \labl{xv}
the number \qq\ turns out to be an integer or half integer in all cases
except for the model of type \Gl\ for which $\qq=5/3$.
The operator product of \hq\ with itself then reads
  \be \hq(z)\, \hq(w) \sim \frac{{\cal N}}{(z - w)^{2}}   , \ee
with
  \be {\cal N} = (\v, \v)\, k + \sumabar (\v, \abar)^2 = (\v, \v) ( k +\gv)
  . \ee
Denote by $\varphi$ a canonically normalized free boson,
satisfying $\ii \partial \varphi(z)\,\ii\partial \varphi (w) \sim
(z - w)^{-2}$. Expressing  \hq\ in terms of $\varphi$,
i.e. $\hq= \sqrt{ {\cal N}}\ii\partial \varphi$, we obtain the \emt
  \be  T =  \frac12  :\ii\partial \varphi \,\ii\partial \varphi : \, =
    \frac1{2 {\cal N}}  : \hq\hq :  \; . \ee
Thus the conformal dimension $\Delta$ of a primary field is
  \be    \Delta = \frac{Q^2}{2 {\cal N}}     , \ee
with $Q$ the \ue-charge of the field, i.e.\ the eigenvalue of \hq.

Thus the \ue\ theory in question is the \cft\ of a free boson
compactified on a torus whose radius is adjusted (or, in other words,
the chiral algebra is enlarged) precisely in such a manner
that the charges are identified modulo $\cal N$.
 \futnote{Thus e.g. $\ue_2$ is the theory for which the extended
algebra is the level one $A_1^{(1)}$ \kma, and $\ue_4\cong{\rm so}(2)_1$.}
 In the sequel we will denote this theory
by $\ue_{\cal N}$. The relevant values of the integer $\cal N$
(as well as the explicit values of the levels of the simple ideals
$h_i$ computed according to \erf a for the
cases of our interest are provided in table \ref{levels}.

For \hsc s it was noticed \cite{sche3} that $\cal N$  is always a
divisor of ${\cal N}_0(g,h)$, where
  \be  {\cal N}_0(g,h)=\ic(g) \cdot \ic(\hat{h})\cdot(k+\gv).  \labl{ic}
Here \ic\ stands for the index of connection (i.e.\ the
number of conjugacy classes, which is equal to the order
of the center $Z$ of the corresponding
universal covering \Lie) of a \lie, and
$\ic(\hat{h})\equiv \prod_i\ic(h_i)$, where $h_i$ are
the simple algebras which appear in the
decomposition $\hat{h} = \mbox{\large{$\oplus$}}_{i} h_{i}$ of $\hat h$
into simple ideals.
In fact, in most cases one even has ${\cal N}={\cal N}_0(g,h)$;
also, by introducing additional identification currents with a
non-trivial component only in the \ue\ part one could use (as
has been done in \cite{sche3}) ${\cal N}_0(g,h)$ in place of
${\cal N}$. For non-\hsc s, however, we encounter two cases,
namely $(\Gl,k)$ and the models $(BA,n,k)$ with $n$ odd,
where the value of $\cal N$ is larger than ${\cal N}_0(g,h)$.

\begin{table}[t] \caption{The values of the levels $k_i$ and of
$\cal N$ for non-hermitian symmetric \cts}
\label{levels} \begin{center} \begin{tabular}{|l|l|} \hline &\\[-3 mm]
\multicolumn{1}{c|}{name} & \multicolumn{1}{|c|}{\coset{g_k\op\sode}
{\bigoplus_i(h_i)^{}_{k_i}\oplus\ue^{}_{\cal N}}{} }
\dhline{3.2}
($BA, n, k$)\, ,\ {\footnotesize $n$ even} &
                \csfull{B_{n}}{n^{2} + n}{A_{n-1}}{k + n -1}{n(k + 2n - 1)}
               \\[1.4mm]
($BA, n, k$)\, ,\ {\footnotesize $n$ odd} &
                \csfull{B_{n}}{n^{2} + n}{A_{n-1}}{k + n -1}{4n(k + 2n - 1)}
               \\[1.4mm]
($BB, 3, k$) & \csfulld{B_{3}}{14}{A_{1}}{2k + 8}{A_{1}}{k +  3}{2(k + 5)}
              \\[1.4mm]
($BB, n, k$)\, ,\ {\footnotesize $n>3$} &
               \csfulld{B_{n}}{8n-10}{B_{n-2}}{k+4}{A_1}{k+2n-3}{2(k+2n-1)}
              \\[1.4mm]
($CC, n, k$) & \csfull{C_{n}}{4n - 2}{C_{n-1}}{k + 1}{2(k + n + 1)}
              \\[1.4mm]
($C3, k $)   & \csfulld{C_{3}}{14}{A_{1}}{k + 2}{A_{1}}{2k + 6}{4(k + 4)}
              \\[1.4mm]
($C4, k $)   & \csfulld{C_{4}}{24}{A_{2}}{2k + 7}{A_{1}}{k + 3}{6(k + 5)}
              \\[1.4mm]
($D4, k $)   & \csfulL{D_{4}}{18}{A_1}{k + 4}{2(k + 6)} \\[1.4mm]
($\dfe, k $) & \csfullD{D_{5}}{30}{A_2}{k + 5}{A_1}{k + 6}{12(k + 8)}
              \\[1.4mm]
($\dfz, k $) & \csfulld{D_{5}}{26}{A_{3}}{k + 4}{A_{1}}{k + 6}{2(k + 8)}
              \\[1.4mm]
($F4, k $)   & \csfull{F_{4}}{30}{C_{3}}{k + 5}{2(k + 9)} \\[1.4mm]
($\Gl, k $)   & \csfull{G_{2}}{10}{A_{1}}{k + 2}{6(k + 4)} \\[1.4mm]
($\Gs, k $)   & \csfull{G_{2}}{10}{A_{1}}{3k + 10}{2(k + 4)}
       \\[1.6 mm] \hline \end{tabular} \end{center} \end{table}

\subsection{Selection rules and field identification}
Our next task is to identify the physical fields of the theories of our
interest. For any \ct, a plausible guess would seem to
be that they are in one to one   correspondence
with the {\em branching functions\/} $b^{\Lambda,{\rm x}}_{\lambda,Q}$
of the embedding. These objects are the coefficient functions
in the decomposition
  \be \Chi_{\Lambda,{\rm x}}^{}(\tau) = \sum_{\lambda} b^{\Lambda,{\rm x}}
  _{\lambda,Q}(\tau)\, \chi_{{}_{\lambda,Q}} (\tau)  \ee
of the product of the characters of $g$
and \sod\ with respect to the characters of $h$.
Here $\Lambda$ and $\lambda$ stand for integrable highest weights of
$g$ and $\hat h$, respectively, and $Q$ for an allowed \ue-charge, while x
denotes an integrable highest
weight of \sod\ at level one, i.e.\ the singlet (0), vector (v),
spinor (s), or conjugate spinor (c) highest weight.

However, this naive ansatz is usually in conflict with the requirement
that the characters of the \ct\ carry a (projective) unitary representation of
the modular group ${\rm PSL}(2,\zet)$. Namely,
on one hand, generically some branching functions turn out
to vanish identically, while the matrix
$S_g S_{h}^*$ that describes the transformation of the branching
functions under the modular transformation $ \tau \mapsto -\frac{1}{\tau}$
has also non-zero elements between vanishing and non-vanishing
branching fanctions (this is an immediate consequence
of the fact that in any \cft\ the modular matrix $S$ obeys
$S_{0 i} \geq S_{0 0} > 0$, and that for any embedding of affine \lie s
the module with highest $h$-weight
zero always appears in the decomposition of the $g$-module with highest
weight zero). On the other hand, typically branching functions for distinct
combinations of highest weights coincide, in such a way that
the matrix $S_g S_{h}^*$
has identical rows and columns and hence cannot be unitary.

One may imagine three distinct sources for the  vanishing of a
branching function. First, usually the matching of
conjugacy classes of $g$- and $h$-modules provides
selection rules. Second, a state that naively
is expected to be a highest weight state may turn out to be a
null state of a Verma module of the affinization of $h$.
And third, it might happen that a given highest weight module $L$
of the reductive subalgebra $h$ does appear
in some module of the affinization of $g$, but that
it gets always combined with other $h$-modules to modules of the
affinization $h^{(1)}$ of $h$ that carry a different highest $h$-weight,
so that (the affine extension of) the
highest weight of $L$ never occurs as a highest
weight of a module of $h^{(1)}$. Although
no general arguments excluding the latter possibility are
known, no example where it is realized has been found so far.
It is therefore common to assume
that this last mentioned possibility never
arises in \cts. Moreover, one usually also assumes that null states
must only be taken into account for $c=0$ \cts.
 \futnote{As already mentioned in the introduction, there exists
one counter example to this assumption, namely \cite{dujo}
the \ct\ $\coset{(A_2)_2}{(A_1)_8}{}$. Note that
the similar \ct\ $\coset{(A_2)_1}{(A_1)_4}{}$ has $c=0$, i.e.\
the underlying embedding is a conformal embedding.}

The correct way to arrive at a modular invariant theory is to
interpret the primary fields of the \ct\ in terms of equivalence
classes of branching functions \cite{gepn8,scya5,hots};
by a slight abuse of terminology,
this prescription is usually referred to
as {\em field identification}. Under the assumptions just
mentioned, the equivalence relation is uniquely determined
by the conjugacy class selection rules.
If all equivalence classes have the same number
of elements, one can simply define a primary field as an equivalence
class of branching functions. Its character is then just any of the
(identical) branching functions of its representatives, and
accordingly the primary field can be
denoted as \pf\Lambda x\lambda Q, where $(\Lambda,{\rm x},\lambda,Q)$\,
is a representative combination of the relevant highest weights
$\Lambda$ of $g$, x of \sod, $\lambda$ of $\hat h$, and $Q$ of \ue.
If, on the other hand, several distinct sizes
of equivalence classes are present,
one has to be more inventive; the
additional manipulations, known as the {\em resolution of
fixed points}, will be addressed  in the next subsection.

Our task is thus to find the relevant selection rules and deduce the
identifications implied by them. This is a
straightforward exercise in group theory, but is still somewhat
involved owing to the non-trivial embedding of $h$ in \sod.
A convenient way to state these selection rules is by means of
simple currents. A simple current $\phi_J$ of a \cft\ is
by definition a primary field whose fusion product with any primary
field $\phi_i$ consists of a single primary field, $\phi_J\star\phi_i
=\phi_{J\star\,i}$\,; the combination
  \be  \qm(\phi_i):= \Delta(\phi_i)+\Delta(\phi_J)-
  \Delta(\phi_{J\star\, i}) \ee
of conformal dimensions is known as the monodromy charge of $\phi_i$
with respect to $\phi_J$. The simple currents of \wzwts\ are all
known \cite{jf15}, and for the \ue\ theories a primary field of
arbitrary charge $Q$ is a simple current.
For any \cft, the subring of the fusion ring that is generated by the
simple currents of the theory is isomorphic to the group ring of
an abelian group $\cal G$ whose
group operation is the one implied by the fusion product. Now denote by
\gs\ the direct product of these groups obtained from
the simple currents of the $g$-, $h_i$- and \ue-parts of the \ct.
It is possible \cite{scya3,scya6}
to characterize the non-vanishing branching functions by the fact that
their monodromy charge with respect to some subgroup of \gs\ vanishes.
We will refer to this subgroup as the {\em identification group\/}
\gid\ of the \ct\ and denote its order by \Sord.  The elements
of \gid\ are usually called identification currents. Their orbits
on the branching functions are just the eqivalence classes
we are looking for. To qualify as an identification current, a simple
current must have integer conformal weight \cite{scya6}
(this allows for a simple
check of our results for the identification currents);
this condition must be met because any identification current
is a representative of the
equivalence class describing the identity primary field,
and conformal weights are constant modulo integers
on each identification orbit.

To begin the description of identification currents for the
theories of our interest, we derive a formula for the order \Sord\
of the identification group of any \nn \ct\ of the form \erf{kasuform}.
 This provides an important check for the completeness of the
selection rules that will be
listed below. Our starting point is the formula \cite{levw}
 \be  \Sord = \left | \frac{L_{g}^{*}}{L_h^\vee} \right | \,. \labl{36}
Here $L$ denotes the root lattice of a reductive algebra, and
$L^\vee$ the corresponding coroot lattice.
The symbol `\,$^{*}$\,' is used to indicate the dual lattice; in
particular $ (L^\vee)^{*} = L^{W}$, where $L^W$ the weight lattice.
Writing the relation \erf{36} in terms of the dual lattices and denoting
the volume of the unit cell by `\vol', we  see that
  \be  \Sord = \left | \frac{(L_h^\vee)^*_{}}{(L_{g}^{*})^*_{}} \right |
     = \left | \frac{L_{h}^{W}}{L_{g}} \right |
     = \frac{\vol (L_{g})}{\vol (L_{h}^{W})}  \,        .   \ee
Since the direction of the \ue\ is orthogonal to $\hat{h}$
in weight space, it follows that
  \be \vol (L_{h}^{W}) = \vol (L_{\hat{h}}^{W})\cdot 1 =
  \vol (L_{\hat{h}}^{W})\ee
and
  \be \vol (L_{g})=  \vol (L_{\hat{h}}) \cdot \qx   ,\ee
where \qx\ is the \ue-charge of the simple root \ax.  Thus
  \be  \Sord = \qx\,\frac{\vol (L_{\hat{h}})}{\vol (L_{\hat{h}}^{W})}
     = \qx\,\left | \frac{L_{\hat{h}}^{W} }{L_{\hat{h}}} \right |
     \equiv \qx\,\ic(\hat{h}) . \labl{ord}
Here $\ic(\hat{h})= \prod_i\ic(h_i)$ as in \erf{ic},
and we made use of the fact that $\ic(h) =  | {L_h^W}/{L_h} | $
for any simple \lie\ $h$.

While the result \erf{ord} is completely general,
the precise form of the group theoretical selection rules
must be determined in a case by case study.
To do so, a rather tedious investigation of the
way $h$ is embedded in $g \oplus \sod$ is necessary. In particular a
careful handling of the embedding of $h$ in \sod\ (best to be
described in an orthogonal basis which corresponds to
the free fermion realization of \sode), which is a special
 \futnote{This is not in conflict with the previously mentioned result
\cite{schW} that \nn symmetry requires regular embeddings. The part of
the embedding that must be regular is $h \hookrightarrow g$ rather than
$h \hookrightarrow \sod$.}
 embedding, is required.
We list in table \ref{ident} our results for the identification
currents $\phi_J$ of all non-hermitian
symmetric \nn \cts\ that can be used in $c=9$ tensor products. We
use the notation $\phi_J\equiv (J^{(g)},J^{({\rm so}(d))}\csp
J^{(h_1)},J^{(h_2)},\ldots,J^{({\rm u}(1))})$. In  the individual
entries, we write \jv\ for the vector simple current, and
\js\ and \jc\ for the spinor and conjugate spinor simple currents,
respectively, of $B$ and $D$ type algebras, while
for $A$ type algebras, $J$ stands for the simple current that
acts as $\mu^i\mapsto \mu^{i+1\,{\rm mod}\,(r+1)}$ on the Dynkin
labels of a $A_r$-weight (this current is
associated with a marginal node
of the Dynkin diagram; it has maximal order, and hence
generates all simple currents of the theory); finally, for the
\ue\ part a field is simply denoted by its \ue-charge $Q$.
Notice that in table \ref{ident} we only give a set of generators of the
group \gid\ rather than all of its elements.

\begin{table}[t] \caption{The identification groups
for non-hermitian symmetric \cts} \label{ident} \label{gid}
\begin{center} \begin{tabular}{|l|c|c|c|} \hline &&&\\[-3.2 mm]
\multicolumn{1}{c|}{name} & \multicolumn{1}{|c|}{\Sord}
& \multicolumn{1}{|c|}{generators of \gid } & \multicolumn{1}{|c|}
{fixed p.} \dhlind{3.7}
($BA, n, k$) % \csfull{B_{n}}{n^{2} + n}{A_{n-1}}{k + n -1}{n(k + 2n - 1)}
              ,\ {\footnotesize $n$ even}
             & $n$ & $(J, 1 \csp J, k + 2n -1) $ & --
               \\[-3.2 mm] &&&\\
($BA, n, k$) % \csfull{B_{n}}{n^{2} + n}{A_{n-1}}{k + n -1}{n(k + 2n - 1)}
              ,\ {\footnotesize $n$ odd}
             & $2n$ & $(J, \jv \csp J, 2(k + 2n -1))$ & --
               \\[-3.2 mm] &&&\\
($BB, n, k$) % \, ,\ {\footnotesize $(n>3)$} &
             % \csfulld{B_{n}}{8n-10}{B_{n-2}}{k+4}{A_1}{k+2n-3}{2(k+2n-1)}
             & 4 & \arraycent{ (J, 1 \csp J, 1, 0) \\[.3 mm] (J, 1 \csp 1, J,
                   \pm (k + 2n -1)) }{1} & \arraycent{++\\[.3 mm]-}{1}
               \\[-3.2 mm] &&&\\
($CC, n, k$) % \csfull{C_{n}}{4n - 2}{C_{n-1}}{k + 1}{2(k + n + 1)}
             & 2 & $ (J, (\jv)^n \csp J, \pm (k + n + 1) ) $ & --
               \\[-3.2 mm] &&&\\
($C3, k $)   % \csfulld{C_{3}}{14}{A_{1}}{k + 2}{A_{1}}{2k + 6}{4(k + 4)}
             & 4 & \arraycent { (J, 1 \csp J, J, 0) \\[.3 mm] (J, \jv
             \csp J, 1, \pm 2(k + 4) ) }{1} & \arraycent{+\\[.3 mm]-}{1}
               \\[-3.2 mm] &&&\\
($C4, k $)   % \csfulld{C_{4}}{24}{A_{2}}{2k + 7}{A_{1}}{k + 3}{6(k + 5)}
             & 6 & $ (J, \jv \csp J, J, -(k+5)) $ & --
               \\[-3.2 mm] &&&\\
($D4, k $)   % \csfull{D_{4}}{18}{A_{1}^{3}}{k + 4}{2(k + 6)}
             & 8 & \arraycent { ( \jc, 1 \csp J, J, 1, 0)  \\[.3 mm]
                 ( \js, 1 \csp J, 1, J, 0) \\[.3 mm]
                %  ( \jv, 1 \csp 1, J, J, 0)    \\[.3 mm]
                    (1, \jv \csp J, J, J, \pm (k+6)) }{1}
                 & \arraycent{+\\[.3 mm] +\\[.3 mm] -}{1} \\[-3.2 mm] &&&\\
($\dfe, k $) % \csfulld{D_{5}}{30}{A_{2}}{k + 6}{A_{1}^{2}}{k + 6}{12(k + 8)}
             &24 & \arraycent { (\js,\jv \csp J,J,1,-(k+8)) \\[.3 mm]
                 (\jv,1 \csp 1, J, J, 0 )) }{1} & \arraycent{-\\[.3 mm]+}{1}
               \\[-3.2 mm] &&&\\
($\dfz, k $) %
             & 8 & \arraycent { (\jv,\jv \csp 1,J,\pm(k+8)) \\[.3 mm]
                 (\js,1 \csp J, J, 0 )) }{1} & \arraycent{-\\[.3 mm]+}{1}
               \\[-3.2 mm] &&&\\
($F4, k $)   % \csfull{F_{4}}{30}{C_{3}}{k + 5}{2(k + 9)}
             & 2 & $ (1, 1 \csp J, \pm (k+9)) $ & -- \\[-3.2 mm] &&&\\
($\Gl, k $)   % \csfull{G_{2}}{10}{A_{1}}{k + 2}{6(k + 4)}
             & 2 & $ (1, \jv  \csp J, \pm 3(k+4)) $ & --
               \\[-3.2 mm] &&&\\
($\Gs, k $)   % \csfull{G_{2}}{10}{A_{1}}{3k + 10}{2(k + 4)}
             & 2 & $ (1, \jv  \csp J, \pm (k+4)) $ & --
\\[1.6 mm] \hline \end{tabular} \end{center} \end{table}

The way in which we arrived at these results is best described by
giving an example. Thus let us have a  look at the \ct\ denoted by
$(C4,k)$. We denote the Dynkin labels of weights of $g = C_4$ by
$\Lambda^i$, $i=1,2,3,4$, of
weights of $h_1 = A_2$ by $\lambda^1$ and $\lambda^2$, of weights
of $h_2 = A_1$ by $\lambda^4$, and the \ue-charge by $Q$. By
analysing the embedding, we find that these numbers must be
related by
  \be  3 \Lambda^1 + 3 \Lambda^3 - 6N + 2 \lambda^1 + 4 \lambda^2 +
  3 \lambda^4 + Q \equiv 0 \bmod 6   ,       \ee
where $6 N$ stands for the sum of six different eigenvalues of the
Cartan generators of so(24), which have integer values in the \NS\
sectors and half integer values in the Ramond sector.
We want to interpret this result as a relation for
monodromy charges, namely
  \be \qm[C_4] +  \qm[{\rm so}(24)] + \qm[A_2] + \qm[A_1] +
  \qm[\ue] \equiv 0 \bmod 1  . \labl{moc}
It is easily checked that $N \bmod \zet$ is the monodromy charge
with respect to the
vector current \jv\ of \sod, and that $Q/p$ is the monodromy
charge with respect to the current  with \ue-charge $- {\cal N}/p$ of
the \ue\ theory. For the $g$ and $h_i$ parts, the identification
currents can also be fixed uniquely, simply because all simple
currents, as well as the associated monodromy charges,
of the corresponding \wzwts\ are known. We then arrive at the
combination
  \be  \phi_J = (J, \jv \csp J, J, -(k+5) )    \ee
of simple currents that has \erf{moc} as its monodromy charge.
This current has order 6. This coincides with the result of
formula (\ref{ord}) for the order of the identification group,
and hence we have already found all identification currents.

\subsection{Fixed points}
If the equivalence classes described in the previous subsection
have different sizes $N$, the identification procedure becomes
more complicated. Note that the maximal size of a class is equal
to the size $N_0=\Sord$ of the equivalence class
of the identity primary field, and that any other allowed size
is a divisor of $N_0$. The equivalence classes of size $N<N_0$
should correspond to $N_0/N$ distinct physical fields \cite{scya5,scya6}.
The required resolution of classes of non-maximal size into primary
fields is problematic because not all necessary
pieces of information are directly supplied by the
embedding; in other words,
the resolution potentially introduces some arbitrariness
in the description of primary fields.
In particular we do not know the characters of the individual
primary fields into which such a class $f$ is resolved. We do know,
however, their sum, since modular invariance imposes the constraint
  \be \sum_i \Chi_{f_i} = \frac{N_0}{N_f}\, \Chi_f,  \labl{Ch}
where $\Chi_f$ denotes the original branching function of the class $f$.

In addition,
fortunately certain sum rules for the modular transformation
matrix $S$ of the full theory can be derived \cite{scya5}.
For brevity, we will refer to a class of non-maximal size as a
{\em fixed point\/} of the field identification.
Now given the naive $S$-matrix element $S_{fg}$ between two fixed points
$f$ and $g$, one can make the ansatz
  \be  \tilde{S}_{f_i\, g_j} = \frac{N_f N_g}{N_0}\, S_{f g}
  + \Gamma^{f g}_{i j}  \ee
for the full $S$-matrix
between different fields $f_i,\, g_i$ into which the fixed points are to
be resolved. The matrix $\Gamma$ introduced here
must be symmetric (with respect to
the double index ($f,i$)), but a priori is otherwise arbitrary.
Modular invariance can be shown to imply the sum rules
  \be \sum_i \Gamma_{i j}^{fg} = 0 = \sum_j \Gamma_{i j}^{fg} . \labl{sumrule}
To find a solution for $\Gamma$ we assume that with respect to the
individual entries of the multi-index $(f,i)\equiv(\Lambda,{\rm x},
\lambda,Q,i)$\, it factorizes as
  \be \Gamma_{i j}^{\Lambda,{\rm x},\lambda,Q;
  \Lambda',{\rm x}',\lambda',Q'} = \Gamma_{(g)}^{\Lambda\,\Lambda'}
  \Gamma_{({\rm so}(d))}^{{\rm x}\,{\rm x}'} \Gamma_{(\hat h)}
  ^{\lambda\,\lambda'} \Gamma_{({\rm u}(1))}^{Q\,Q'} P_{i j}, \labl{fa}
where
  \be P_{i j} = \delta_{i j} - \frac{N_f}{N_0}\,.  \labl{pij}
Since in all cases of our interest the fixed points $f$ have order
$N_0/N_f=2$ and must therfore be resolved into two fields, the fact
that \erf{pij} can be factored out is an immediate consequence
of the sum rules (\ref{sumrule}).
Following \cite{scya6}, with the factorization assumption \erf{fa}
we can identify in all cases a so-called
fixed point \cft, whose characters can be added to the branching functions
to get the full collection of primary fields;
these characters are nothing but the summands $\Chi_{f_i}(\tau)$
in the decomposition \erf{Ch} above.

This procedure of fixed point resolution is certainly quite
important, because it is only after having
accomplished this task that we really deal with a well-defined
\cft\ (it is even unknown whether the prescription works for an
arbitrary \ct, and whether the \cft\ it provides is unique).
However, it is not difficult to see that some important
quantities we will be
interested in, namely the number of generations and anti-generations
in a four-dimensional string compactification, can be obtained in our
case without a detailed knowledge of the resolution procedure
(see also the comments in the following section).

In the third column of
table \ref{ident} we marked whether identification fixed points
occur in the theories in question. The following notation is used:
`--' indicates that fixed points never occur in the corresponding theory;
`+' means that fixed points can occur, but not at any of
the levels that are relevant for $c=9$ tensor products (this typically happens
when we are only interested in low levels where the associated outer
automorphisms of $g$ act freely on the integrable representations of $g$);
finally `++' is used to indicate that fixed points occur and have to be
resolved. Note that an identification current can possess a fixed
point only if it has vanishing \ue-charge.

\subsection{Modular invariants}
It should be noted that the discussion of field identification in
the previous subsections refers only to  one chiral half of
the \cft. For the full theory, one has to use all fields
as identification currents, i.e.\ as representatives of the
identity primary field, that have non-vanishing branching functions
and are identification currents with
respect to both the holomorphic and the anti-holomorphic part.
For example, for the \nn minimal models this prescription implies
the presence of left-right asymmetric identification currents if
the $D_{\rm even}$, $E_6$, or $E_8$ type invariants of the
associated $A_1$ \wzwt\ are chosen.

For the \nn theories of our present interest, we will confine
ourselves to analyse only the situation where the diagonal modular
invariants of $g$, $h$ and \sod\ are used. As a consequence,
the identification currents are just the left-right symmetric
version of the chiral currents listed in table \ref{gid}.
The extension to any known non-diagonal modular invariant
is immediate; note however that the classification of modular
invariants of simple Lie algebras other than $A_1$
(and of their tensor products) is far from being complete.

\sect{Chiral ring and \pop s}
In this section we present some  results for quantities relevant
to string compactification. We have computed the quantities which
are the most relevant ones for the phenomenological aspects,
namely the number of (anti-)generations for a compactification
of the heterotic string to four space-time dimensions.
To obtain these numbers, we need some information on the
collection of {\em chiral primary fields\/} of the theory;
these fields are by definition
those primary fields which satisfy $\qsuco = \Delta /2$.
They generate the {\em chiral ring\/} \cite{levw} of the theory; this
is a finite-dimensional nilpotent ring $\cal R$ whose product is
the naive operator product $\lim_{z\rightarrow w}\phi(z)\phi'(w)$.

For the models under consideration, it is in fact easier to work with
the ground states of the \R sector, which owing to spectral flow
\cite{levw} provide equivalent information on the theory. Namely,
the chiral primary fields (with superconformal charge
 \futnote{In the literature sometimes a normalization is chosen
where $\onehalf q_{\rm suco}$ is the superconformal charge.}
 \qsuco)
are via spectral flow in one to one correspondence with \rgs s
(with superconformal charge $\qsuco-c/6$).
In all \nn coset models of the form (\ref{kasuform}) we can identify
the simple current in the \R sector which generates the flow;
it is the unique \rgs\ with highest superconformal charge, which has been
termed {\em spinor current\/} in \cite{sche3}. It is easily seen
that one representative of the spinor current is the field
  \be  S=\pf0s0{Q_s} , \labl S
with
  \be  Q_s = ( \v , \rhog - \rhoh). \ee
Here $\rho_g=\sum_i\la i$ and \rhoh\ are the Weyl vectors, i.e.\
half the sum of positive roots, of $g$ and of $\hat h$, respectively.

The information on the multiplicities of chiral states with a given
superconformal charge is encoded in the {\em \pop\/} \cite{levw},
which can be defined as a trace over the chiral ring ${\cal R}$,
  \be  P(t,\bar t\,) := \mbox{Tr}^{}_{\cal R}\,
  t^{J_0} \bar{t}^{\bar{J}_0}.  \ee
Here $J_0$ denotes the generator of the superconformal \ue, and
the barred quantities refer to the second chiral half of the theory.
In the sequel we will only consider the left-right symmetric
diagonal modular invariant; correspondingly we can
restrict ourselves to one chiral half and replace $t \bar{t}$
for the sake of simplicity by $t$.

\subsection{\rgs s}
To determine the ground states of the Ramond sector one can use a simple
formula for the $g$- and $h$-weights of these states which can be derived
\cite{levw} by means of an index argument.  The advantage of this
formula is twofold. First, in coset models it is usually difficult to
calculate the integer part of the conformal weight $\Delta$ of a primary
field; for \rgs s (which all have $\Delta=c/24$),
however, the index argument makes it possible to identify the state
without having to evaluate a formula for $\Delta$.
Second, the formula automatically takes care of possibly arising null
states; again, this is a rather delicate issue in the general case.
 \futnote{Compare the remark about $E_6$ singlets in section \ref{44}
below.}

Denote by \wg\ the Weyl group of $g$, by $|\wg|$ its order,
and by \wh\ and $|\wh|$ the analogous quantities for $\hat h$.
For any integrable $g$-weight $\Lambda$, the recipe of \cite{levw} provides
$|W_g|/|W_h|$ \rgs s.  The $h$-weight \lafull\
of each of these \rgs s is related to its $g$-weight by
  \be \lafull = w ( \Lambda + \rho_g) - \rho_h \,. \labl{51}
Here the weight $\lafull\equiv(\lambda,Q)$ incorporates both the
weight $\lambda$ of the semi-simple part $\hat h$ of $h$
and the \ue-charge $Q$. Also, the map $w$ in \erf{51}
is the representative of any class of the coset \wgh\
possessing the property that \lafull\
is a dominant integral highest weight of $h$; each class of \wgh\
contains a unique representative $w$ satisfying this requirement
\cite{kost3}. If sign$(w)=1$, then the highest weight of \sod\
that is associated to
$\lafull$ is the spinor (s), while for sign$(w)=-1$, it is the
conjugate spinor (c). Note that $\rhog-\rhoh\propto\lx$, the
constant of proportionality being $\sum_j(\la j,\lx)/(\lx,\lx)$
as can be deduced from $(\rhoh,\rhog-\rhoh)=0$.

To implement the formula \erf{51} on a computer, it is convenient not
to start with the weights of $g$, but rather to scan all dominant
weights of $h$ that are allowed by the selection rules. For each
such weight $\lafull$ one
determines the unique dominant integral $g$-weight which lies on the
same \wg-orbit as $\lafull + \rho_h$
(if this $g$-weight is not integrable at the relevant level of the affine
algebra $g^{(1)}$, then the corresponding state has to be rejected).
To do so, one only has to
know the action of the fundamental reflections $w_i\in\wg$
(see e.g.\ \cite{fuva3}).
This method has the advantage that one needs not know the
whole \wg-orbit of a highest $g$-weight which, especially for large
rank algebras, would require a lot of memory.

\subsection{\pop s}
Having found the \rgs s, we can proceed to compute the \pop\
of an \nn \ct. To do so, we also need the superconformal
charge of the \rgs s. This charge is given by \cite{kasu2}
  \be  \qsuco= \sumabar \labar - \frac{\qq Q}{k + \gv} \,. \labl{qs1}
Here $\qq=\sqrt{(\Q,\Q)/(\v,\v)}=\sqrt{(k+\gv)(\Q,\Q)/{\cal N}}$
is the number defined by \erf{xv}, $Q$ is the \ue-charge of the \rgs,
and $\labar\in\{\onehalf,-\onehalf\}$ are the components
of its \sod-weight in the orthogonal basis.
Unfortunately the index argument \cite{levw} leading to \erf{51}
does not provide the full weight $\tilde\Lambda$,
but only yields the information whether it is a weight of the
the spinor or of the conjugate spinor module of \sod, or in
other words, only the value of $\sumabaR\labar$ modulo 2.
To translate \erf{qs1} into a more convenient formula, we proceed as
follows.
 \futnote{An analogous result has been obtained in \cite{lewa2} for
simply laced \hsc s at level one, and in \cite{ekmy}
for all \hsc s in their free field realization.}
 Denote by $\delgp$, $\delgm$, and
$\Delta^g_{}$ the sets of positive roots, of negative roots,
and of all roots, respectively, of the \lie\ $g$,
and by $\delhpm$, $\Delta^h_{}$ the corresponding quantities for $\hat h$.
For an arbitrary element $w$ of the Weyl group \wg\ define
  \be \delwpm := \{ \alpha \in \delg \mid w^{-1} (\alpha) \in \delgpm\}. \ee
For any $w\in\wg$, $\delg$ is the disjoint union of $\delwp$ and $\delwm$.
We can express the image of the Weyl vector \rhog\ under $w$ as
\be    w( \rhog) = \onehalf\,\mbox{\large[} \sum_{\alpha \in \delwp} \alpha
          - \sum_{\alpha \in \delwm} \alpha\,\mbox{\large]} \ ,   \ee
as is easily verified by applying $w^{-1}$ to both sides of the equation.

Given a subalgebra $h$ of $g$, we call $w \in\wg$ {\em $h$-positive\/}
\cite{lewa2}, iff
  \be   \delhp \subset \delwp \,  .  \ee
We claim that in order to compute $\sumabaR\labar$,
we only need to identify the $h$-positive
representative  $w$ of the coset \wgh\ that appears in \erf{51}, and that
the components \labar\ of the \sod-weight $\tilde\Lambda$ are given by
  \be  \labar=\labarw:= \left\{ \begin{array}{rll} \onehalf & \mbox{if} &
  \abar \in \delwp,  \\[1 mm] -\onehalf & {\rm if} & \abar\in\delwm
  . \end{array}\right. \labl{pm}
This can be seen as follows.
Let $\alpha$ be an arbitrary element of $\delhp$. For any
highest $h$-weight $\lafull$ we have $( \lafull + \rhoh, \alpha) > 0$;
as a consequence,
\be    0 < (\lafull + \rhoh, \alpha) = (w(\Lambda + \rhog), \alpha)
         = (\Lambda + \rhog, w^{-1} (\alpha)) .   \ee
This shows that $w^{-1} (\alpha) \in \delgp$, or in other words, that
$\delhp \subset\delwp$.
Now the general form of the Cartan currents of $\hat h$ reads
  \be   H^i_{\hat h}= H^i_g + \sumabar \abar^i\,\nabar \,. \ee
As a consequence, under the embedding $ h \hookrightarrow g\op\sod$
the state with weight $( w(\Lambda), \labarw)$ branches to
  \be \begin{array}{ll} \lafull\! &
   = \;w(\Lambda) + \onehalf {\displaystyle\sum_{\bar\alpha \in \delwp}}
        \abar - \onehalf {\displaystyle\sum_{\bar\alpha\in \delwm}} \abar
   \\ {}&{} \\[-2 mm]
     & = \; w(\Lambda) + w(\rhog) - \onehalf\! {\displaystyle\sum_{\alpha
     \in\delwp\cap\delhp}}\!\! \alpha\, + \onehalf\! {\displaystyle\sum_
     {\alpha\in\delwm\cap\delhp}}\!\! \alpha \;  . \end{array} \ee
This reduces to $w( \Lambda + \rhog) - \rhoh$, i.e.\ yields
the correct result \erf{51}, iff $w$ is $h$-positive. Note that the
weight $(w(\Lambda), \labarw)$ is always present in the weight system
of the $g\op\sod$\,-module with highest weight $(\Lambda,{\rm s})$
or $(\Lambda,{\rm c})$, because the Weyl group orbit
of any weight of a highest weight module with dominant integral
highest weight is contained in the weight system of the module.

Inserting our result \erf{pm} into the formula \erf{qs1} for the
superconformal charge \qsuco, we obtain
  \be   \qsuco = \onehalf\, \mbox{\large(}{ | \delwp \cap \Delta_{+}| -
    |\delwm\cap\Delta_+| }\mbox{\large)} - \frac{\qq Q}{k+\gv} \, .  \ee
To simplify this formula further, we recall that
the length $l(w)$ of a Weyl group element $w$, which is
defined as the minimal number of fundamental reflections
needed to obtain $w$, obeys \cite[sect.\,1.7]{HUmp2}
  \be l(w)  = | \delwm \cap \Delta_{+} |    .\labl{lw}
Using  the identity
  \be    |\delwp \cap \Delta_+ | + |\delwm \cap \Delta_+| = d ,\labl{dd}
we finally obtain
  \be   \qsuco = \onehalf d - l(w) - \frac{\qq Q}{k +\gv} \, .  \labl{qs2}
The length of the relevant elements of \wgh\ can be obtained
conveniently via the so-called
Hasse diagram of the embedding \hing\ (for some details, see
the Appendix), and hence the formula \erf{qs2} is easily
implemented in a computer program.
For the spinor current \erf S, one has $w={\sl id}$ so that
\erf{qs2} reduces to
  \be  \qsuco(S)=\onehalf d - \frac{(\Q,\rhog-\rhoh)}{k +\gv}
  =\frac c6 \,, \ee
where the last equality follows with \erf{26} and the
strange formula.

We are now in a position to compute the \pop s of the theories
listed in section 2. For notational simplicity, we will present the
\pop s in the form $P(t^\de)$, with $\de$ the smallest positive integer
for which all values of $\de\qsuco$ of chiral primary fields are
integers. We find that for the three series
$(BB,m+2,1)$, $(CC,2,2m+1)$, and $(CC,2m+2,1)$ with $m\in\zetpluso$,
the \pop s are given by a common formula, namely $\de=m+2$ and
  \be  P(t^{m+2})=\sum_{j=0}^m (j+1)\,(t^j+t^{3m+2-j})+(3m+4)
  \sum_{j=m+1}^{2m+1} t^j. \labl{bc}
The \pop s of the remaining models are listed in table \ref{pop}.

To conclude this subsection, we remark that the resolution
of fixed points does not alter the number of \rgs s. In
other words, independently of its length each identification orbit
that contains a representative satisfying \erf{51}
provides exactly one \rgs\ \cite{sche3}.

\begin{table}[t] \caption{\pop s for non-hermitian symmetric \cts}
\label{pop} \begin{center} \begin{tabular}{|l|r|l|} \hline &&\\[-3 mm]
\multicolumn{1}{|c|}{name} & \multicolumn{1}{c|}{$\de$}
& \multicolumn{1}{c|}{$P(t^\de)$} \dhlinz{3.7}

 $(BA ,  3,  1)$ & 4 &
 $  % P(t^{  4})
  1 + 3 \, t^{  2} + 4 \, t^{  3} + 3 \, t^{  4} + t^{  6} $ \\[.9 mm]

 $(BA ,  3,  2)$ & 14 &
 $  % P(t^{  14})
  1 + t^{  6} + t^{  8} + t^{  9} + 2 \, t^{  10}
+ t^{  11} + 3 \, t^{  12} + t^{  13} +
2 \, (\, t^{  14} + t^{  15} + t^{  16} \,)
$ \\ && $ \qquad
+ t^{  17} + 3 \, t^{  18} + t^{  19} + 2 \, t^{  20}
+ t^{  21} + t^{  22} + t^{  24} + t^{  30} $ \\[.9 mm]

 $(BA ,  3,  4)$ & 6 &
 $  % P(t^{  6})
  1 + (\, t^{  2} + t^{  3} \,) + 3 \, t^{  4} + 2 \,
t^{  5} + 9 \, t^{  6} + 7 \, t^{  7} + 14 \, t^{  8}
+ 12 \, t^{  9} $ \\ && $ \qquad
+ 14 \, t^{  10} + 7 \, t^{  11} + 9 \, t^{  12} + 2 \, t^{  13}
+ 3 \, t^{  14} + t^{  15} + t^{  16} + t^{  18} $ \\[.9 mm]

 $(BA ,  4,  1)$ , % & 2 &
  % $  % P(t^{  2})
   % 1 + 4 \, t + 14 \, t^{  2} + 4 \, t^{  3} + t^{  4} $ \\[.9 mm]

 $ (C3 ,  1) $ , %  & 2 &
  % $  % P(t^{  2})
   % 1 + 4 \, t + 14 \, t^{  2} + 4 \, t^{  3} + t^{  4} $ \\[.9 mm]

 $ (\Gs,  2) $  & 2 &
 $  % P(t^{  2})
  1 + 4 \, t + 14 \, t^{  2} + 4 \, t^{  3} + t^{  4} $ \\[.9 mm]

 $(BA ,  5,  1)$ & 4 &
 $  % P(t^{  4})
  1 + 5 \, t^{  2} + 10 \, t^{  4} + 16 \, t^{  5} + 10 \,
t^{  6} + 5 \, t^{  8} + t^{  10} $ \\[.9 mm]

 $(BA ,  6,  1)$ & 2 &
 $  % P(t^{  2})
  1 + 6 \, t + 15 \, t^{  2} + 52 \, t^{  3} + 15 \,
t^{  4} + 6 \, t^{  5} + t^{  6} $ \\[.9 mm]

 $(BB ,  3,  3)$ & 2 &
   $  % P(t^{  2})
    1 + 4 \, t + 17 \, t^2 + 40 \, t^3 + 17\,t^4
    + 4\,t^5 + t^6 $ \\[.9 mm]

 $(BB ,  4,  2)$ & 3 &
   $  % P(t^{  3})
    1 + 2 \, t + 8 \, t^2 + 14 \, t^3 + 35\,t^4 + 35\,t^5 + 14\,t^6
    + 8\,t^7 + 2\,t^8 + t^9 $ \\[.9 mm]

 $(CC ,  2,  2)$ , % & 5 &
  % $  % P(t^{  5})
   % 1 + 2 \, t^{  2} + 3 \, (\, t^{  3} + t^{  4} \,) +
   % 2 \, t^{  5} + t^{  7} $ \\[.9 mm]

 $ (CC ,  3,  1) $ , % & 5 &
  % $  % P(t^{  5})
   % 1 + 2 \, t^{  2} + 3 \, (\, t^{  3} + t^{  4} \,)
 % + 2 \, t^{  5} + t^{  7} $ \\[.9 mm]

 $ (\Gs,  1) $  & 5 &
 $  % P(t^{  5})
  1 + 2 \, t^{  2} + 3 \, (\, t^{  3} + t^{  4} \,)
+ 2 \, t^{  5} + t^{  7} $ \\[.9 mm]

 $(CC ,  2,  4)$ , % & 7 &
  % $  % P(t^{  7})
   % 1 + 2 \, t^{  2} + 3 \, t^{  4} + 5 \, t^{  5} + 4 \,
 % (\, t^{  6} + t^{  7} \,) + 5 \, t^{  8} + 3 \, t^{  9} + 2 \, t^{  11}
 % + t^{  13} $ \\[.9 mm]

 $ (CC ,  5,  1) $  & 7 &
 $  % P(t^{  7})
  1 + 2 \, t^{  2} + 3 \, t^{  4} + 5 \, t^{  5} + 4 \, (\,
t^{  6} + t^{  7} \,) + 5 \, t^{  8} + 3 \, t^{  9}
+ 2 \, t^{  11} + t^{  13} $ \\[.9 mm]

 $(CC ,  2,  6)$ , % & 9 &
 % $  % P(t^{  9})
  % 1 + 2 \, t^{  2} + 3 \, t^{  4} + 4 \, t^{  6} + 7 \,
% t^{  7} + 5 \, t^{  8} + 6 \, (\, t^{  9} + t^{  10} \,) + 5 \, t^{  11}
% + 7 \, t^{  12} + 4 \, t^{  13} + 3 \, t^{  15} + 2 \, t^{  17} + t^{  19} $
% \\[.9 mm]

 $ (CC ,  7,  1) $  & 9 &
 $  % P(t^{  9})
  1 + 2 \, t^{  2} + 3 \, t^{  4} + 4 \, t^{  6} + 7 \,
t^{  7} + 5 \, t^{  8} + 6 \, (\, t^{  9} + t^{  10} \,) $
\\ && $ \qquad + 5 \, t^{  11}
+ 7 \, t^{  12} + 4 \, t^{  13} + 3 \, t^{  15} + 2 \, t^{  17} + t^{  19} $
\\[.9 mm]

 $ (CC ,  3,  2) $ ,  & 2 &
   $  % P(t^{  2})
   1 + 6 \, t + 16 \, t^{  2} + 6 \, t^{  3} + t^{  4} $ \\[.9 mm]

 $ (CC ,  3,  5) $ , % & 3 &
  % $  % P(t^{  3})
   % 1 + 3 \, t + 12 \, t^{  2} + 20 \, t^{  3} + 48 \, (\,
 % t^{  4} + t^{  5} \,) + 20 \, t^{  6} + 12 \, t^{  7} + 3 \, t^{  8} +
 % t^{  9} $ \\[.9 mm]

 $ (CC ,  6,  2) $  & 3 &
 $  % P(t^{  3})
  1 + 3 \, t + 12 \, t^{  2} + 20 \, t^{  3} + 48 \, (\, t^{  4}
+ t^{  5} \,) + 20 \, t^{  6} + 12 \, t^{  7} + 3 \, t^{  8} + t^{  9} $
\\[.9 mm]

 $ (CC ,  4,  3) $  & 2 &
 $  % P(t^{  2})
  1 + 8 \, t + 29 \, t^{  2} + 64 \, t^{  3} + 29 \, t^{  4}
+ 8 \, t^{  5} + t^{  6} $ \\[.9 mm]

 $ (C4 ,  1) $  & 2 &
 $  % P(t^{  2})
  1 + 4 \, t + 15 \, t^{  2} + 40 \, t^{  3} + 15 \, t^{  4}
+ 4 \, t^{  5} + t^{  6} $ \\[.9 mm]

 $ (D4 ,  1) $  & 7 &
 $  % P(t^{  7})
  1 + t^{  2} + 3 \, t^{  4} + 4 \, t^{  5} + 3 \, (\, t^{  6}
+ t^{  7} \,) + 4 \, t^{  8} + 3 \, t^{  9} + t^{  11} + t^{  13} $ \\[.9 mm]

 $ (\dfe, 1) $ & 3 &
 $ % P(t^{  3})=
 1 + t + 4 \, t^{  2} + 12 \, t^{  3} + 22 \, (\, t^{  4} + t^{  5} \,)
+ 12 \, t^{  6} + 4 \, t^{  7} + t^{  8} + t^{  9} $ \\[.9 mm]

 $ (\dfz, 1) $ & 9 &
 $ % P(t^{  9})=
 1 + t^{  2} + 2 \, t^{  4} + 3 \, t^{  6} + 5 \, t^{  7} + 4 \, (\,
t^{  8} + t^{  9} + t^{  10} + t^{  11} \,) $
\\ && $ \qquad + 5 \, t^{  12} + 3 \,
t^{  13} + 2 \, t^{  15} + t^{  17} + t^{  19} $ \\[.9 mm]

 $ (F4 , 1) $ & 5 &
 $ % P(t^{  5})=
 1 + t + 2 \, (\, t^{  2} + t^{  3} \,) + 9 \, (\, t^{  4}
+ t^{  5} + t^{  6} + t^{  7} \,) $
\\ && $ \qquad + 2 \, (\, t^{  8} + t^{  9} \,) + t^{  10} + t^{  11} $ \\[.9
mm]

 $ (\Gl,  1) $  & 3 &
 $  % P(t^{  3})
  1 + 5 \, (\, t^{  2} + t^{  3} \,) + t^{  5} $ \\[.9 mm]

 $ (\Gl,  2) $  & 18 &
 $  % P(t^{  18})
  1 + t^{  10} + t^{  12} + t^{  13} + t^{  14} + t^{  15}
+ 2 \, t^{  16} + t^{  17} + t^{  18} + t^{  19} + 2 \, t^{  20}
$ \\ && $ \qquad + t^{  21}
+ t^{  22} + t^{  23} + 2 \, t^{  24} + t^{  25} + t^{  26} + t^{  27}
+ t^{  28} + t^{  30} + t^{  40} $ \\[.9 mm]

 $ (\Gs,  5) $  & 3 &
 $  % P(t^{  3})
  1 + t + 4 \, t^{  2} + 8 \, t^{  3} + 22 \, (\, t^{  4}
+ t^{  5} \,) + 8 \, t^{  6} + 4 \, t^{  7} + t^{  8} + t^{  9} $
\\[1.6 mm] \hline \end{tabular} \end{center} \end{table}
\clearpg

\subsection{Charge conjugation}
{}From \erf{bc} and the results in table \ref{pop}
one can read off that the superconformal charges of chiral
primary fields lie between zero and $c/3$, as it must be.
One also notes that according to the results the \pop s obey
  \be  P(t)=t^{c/3}\,P(t^{-1}) . \ee
In terms of the \R sector this means that the collection of \rgs s
is symmetric with respect to the charge conjugation
$\qsuco\mapsto-\qsuco$.

In fact, using the formul\ae\ \erf{51} and \erf{qs2} it is possible to
show that this is a generic feature of all \nn \cts\ of the form
(\ref{kasuform}).
To show this, consider along with an arbitrary \rgs\
$\pF\Lambda{\rm x}{\laful}\equiv \pf\Lambda{\rm x}\lambda Q $
also the field represented by \pF{\Lambda^+}{\rm x'}{\lafulp},
with $\Lambda^+_{}$, x$'$ and $\lafulp$ defined as follows.
As before, x stands for either the spinor or conjugate spinor,
and we define x$'$ to
be equal to x if $d$ is even, and to belong to the opposite
conjugacy class if $d$ is odd. Moreover,
  \begin{eqnarray} \Lambda^+_{} &:=& - \wmax{g} (\Lambda)  , \\ [1.5 mm]
  \laful^+ &:=& - \wmax{h} (\lafull) \label{l+} \end{eqnarray}
(recall that \wmx, denoting the longest element of a Weyl group $W$,
acts as the negative of the conjugation in the
representation ring of a \lie). In the definition \erf{l+},
\wmax h\ is to be considered as an element of the Weyl group
$W_g$. As a consequence, \wmax h\ acts on the $\hat h$-weights
like the
usual conjugation of weights and maps $Q$ to $-Q$. Namely,
by virtue of (\ref{hpro}) each fundamental Weyl reflection of
$W_h$, and thus any element of $W_h$, acts on \v\ as the identity.

Using the identities $ \rhog = - \wmax{g}(\rhog)$ and
 $ \rhoh = - \wmax{h}(\rhoh)$,
we see that the highest weight $\laful^+ + \rhoh $ of $h$
can be written as
  \be  \begin{array}{ll} \lafulp + \rhoh
  & =\, - \wmax{h} (\lafull + \rhoh )
  \,=\, - \wmax{h} w (\Lambda + \rhog)  \\[2.4 mm]
  & =\, w^+ (\Lambda^+ + \rhog ) \,, \end{array}\ee
where
  \be   w^+ := \wmax{h} w\, \wmax{g} \,, \ee
and where $w$ is the Weyl group element introduced in
(\ref{51}). To calculate the
sign of $w^+$, which determines the \sod\ conjugacy class, we observe
(by inspection) that for all simple \lie s $g$ the relation
  \be    \sign(\wmax g)= (-1) ^{n_{+}}   \ee
is satisfied,
where $n_{+}=|\delgp|$ is the number of positive $g$-roots. Therefore
  \be \sign(w^+)= \sign (\wmax h)\, \sign (\wmax g)\, \sign (w)
    = (-1)^{(\dim g -\dim h )/ 2}\, \sign(w), \ee
and (\ref{51}) now clearly implies
that the state \pF{\Lambda^+}{x'}{\lafulp} is again
a \rgs.  Also note that as a by-product we proved that along
with $w$ also $w^+$ is $h$-positive.

So far we have seen that the set of \rgs s is symmetric in the
\ue-charge. The symmetry in  the superconformal charge then follows
from \erf{qs2} together with the identity
  \be  l(w) + l(w^+ ) =  d  . \labl{30}
This relation arises as follows. Let \abar\
be an arbitrary root in $\Delta_+$.
Then either $ w^{-1} (\abar ) \in \delgm $ or
$( w^+)^{-1}  (\abar ) \in \delgm $, because the map
$w\mapsto w^+$ swaps exactly from negative to positive roots of
$g\setminus h$. Thus $\Delta _+$ is the
disjoint union of $\Delta_+ \cap \delwm $ and $\Delta_+ \cap \Delta_-
^{[w^+]}$, which by \erf{lw} and \erf{dd} proves the assertion.

Let us also note that the unique \rgs\ with minimal superconformal
charge $\qsuco=-c/6$ (which via spectral flow corresponds
to the identity primary field) is obtained by applying the
above prescription to the spinor current \erf S, and hence is
given by \pf0{{\rm x}'}0{-Q_s}. For this field the relevant
$h$-positive Weyl group element is $w=\wmax h\wmax g$
so that \erf{30} implies
  \be  l(\wmax h \wmax g )=d \,, \labl{whg}
while by setting $\Lambda=\lambda=0$ in \erf{51}, one obtains
$\rhoh=\onehalf\,(\rhog+\wmax h \wmax g (\rhog))
=\onehalf\,(\rhog-\wmax h (\rhog)).$
Since the $h$-positive representative \wmax{g/h}\
of \wgh\ with largest length $d$ is unique
\cite{HUmp2}, \erf{whg} shows that this representative is given by
  \be  \wmax{g/h}=\wmax h \wmax g \,. \ee

\subsection{Extended \pop s and string theory spectra} \label{44}
Knowing the exact form of the \rgs s, we are in a position
to calculate the massless spectrum of the string theory that
employs an \nn coset model as its inner part, or more
precisely, the numbers \ngen\ of `generations' and \nagen\ of
`anti-generations' which carry the two inequivalent
27-dimensional representations of the $E_6$ part of the
space-time gauge group of the string theory.
One possibility to find these numbers (and, in addition, the
number $N_1$ of $E_6$ singlets) is the `method of beta vectors'
that was introduced \cite{gepn3} in the context of \nn
minimal models. In practise, this is not the most convenient approach,
as the dimensionality and structure of the lattice spanned by the beta
vectors depends strongly on the algebras involved,
so that one would be forced into a lengthy case by case analysis.
(However, for the calculation of the number of massless states
carrying the singlet representation of the space-time gauge group $E_6$,
the method of beta vectors is still the only known algorithm.
Unfortunately the knowledge of the \rgs s is {\em not\/}
sufficient to get the singlets. Now while for \rgs s the
correct treatment of null states is already implemented
through \erf{51}, for general \nn \cts\ the presence of null states
makes the determination of the singlets a hard problem.
In fact, the singlet numbers have
so far not been determined for (tensor products of) \nn
coset models other than the minimal ones. For the latter theories,
the representation theory of the \nn algebra
gives a good handle on null states.)

An alternative algorithm is provided by the {\em
\epop\/} $\cal P$ that was introduced in \cite{sche3}. This
polynomial depends on the variable $t$ of the ordinary \pop\
and on an additional variable $x$ which keeps track of the
intersections of the orbits of the spinor current $S$
with the set of \rgs s.
To determine the exact form of the extended \pop\ is a somewhat
tricky issue, as in fact we only know some specific
representatives of the fields which are \rgs s
(the formula (\ref{51}) does not provide all members of an
eqivalence class), while for the calculation of \ptx\
in principle all representatives are required. Fortunately,
one can show that the following procedure yields the
full result. Take a single representative for each \rgs, and
act on it with all even powers of all representatives of $S$
that  have $g$-weight $\Lambda = 0$. This is sufficient because of the
fact, proven in the appendix of \cite{levw}, that for any representative
$R$ of a \rgs\ there exists at least one representative
$R'$ that belongs to the set
obtained via (\ref{51}) and that has the same $g$-weight as $R$.
If any of the states so obtained is either
a \rgs\ or a superpartner of a \rgs, there is a
corresponding contribution
  \be  \pm\,t^{q_{\rm suco}(R)+c/6}\,x^m \ee
to the extended \pop, with $m$ the power of $S$ that has been
applied, and with the sign being positive for the case of a \rgs\
and negative for the case of the superpartner of a \rgs,
respectively. Here by `superpartner' of a state we mean the state
that is obtained by taking the fusion product with the simple
current \pf0v00\ which is
the generator of the world-sheet supersymmetry.
 \futnote{Note that a primary field and the field related to it
by world-sheet supersymmetry are to be treated as distinct primary fields.
In particular \pf0v00\ is itself a primary field.}
 Note that for large enough power $M$ the \rgs\ itself
or its superpartner is reproduced, so that in fact one has
an infinite power series in $x$, which however is periodic such
that it can be factored into a polynomial times $(1\mp x^M)^{-1}$.

If in \ptx\ the
highest (and, due to charge conjugation invariance proven
in section 4.3, also the lowest) power in $t$ gets multiplied with
more than two distinct powers of $x$, then additional gravitinos that lead
to extended space-time supersymmetry (respectively, additional gauge
bosons, yielding an extension of the space-time gauge group $E_6$
to $E_7$ or $E_8$) are present \cite{sche3}.
In the tables below we have marked all models
where this happens by an asterisk on the net generation number.
Note that in the tables we display the number
of $E_6$ multiplets even if the gauge group gets extended.
(All models of this type that appear in our list
describe in fact string propagation on the manifold
$K\!{\sl3}\times T^2$, and hence have $\ngen=\nagen=21$.
The number $N_{56}^{}$ of the associated $E_7$ multiplets is
in these cases $N_{56}^{}=\ngen-1=20$,
as one generation-antigeneration pair becomes part of the
gauge boson multiplet.)

If the gauge symmetry is not extended, then
as argued in \cite{sche3} it is straightforward to read off the
numbers \ngen\ and \nagen\ from the extended \pop\ of a \cc9
theory. Namely, if $\cal P$ is written as
  \be  \ptx=\sum_i \sum_{m=0}^\infty a^{(q_i)}_m\,t^{q_i}_{}
  x^{2m} , \ee
then
  \be \ngen+\nagen=\sum_{m=0}^{\mo/2-1} |a^{(1)}_m| \,, \qquad
      \ngen-\nagen=\sum_{m=0}^{\mo/2-1} (-1)^m a^{(1)}_m \,. \labl{ep}
Here $\mo$ denotes the smallest positive integer such that
the $(2\mo+1)$st power of the spinor current is
either equal to the spinor current itself or to its superpartner.

As an illustration, we present one example of an extended \pop, namely
for the theory $(\Gl,2)$. This has the
(somewhat atypical) property that to some powers of $t$ other than
the highest and the lowest ones there are associated
more than two different powers of $x$. The `polynomial' reads
  \be   \begin{array}{ll}  {\cal P}(t^{18},x) \,=&  \!\!\!
       \{     \, (1 + x^{ 2})
  + \, t^{10} \, (1 + x^{10}) %\\[1.3 mm]
  + \, t^{12} \, (1 + x^{ 8} + x^{18} + x^{26}) %\\[1.3 mm]
  + \, t^{13} \, (1 - x^{16})  \\[1.3 mm]
& + \, t^{14} \, (1 + x^{ 6} - x^{12} - x^{30}) %\\
  + \, t^{15} \, (1 - x^{32}) %\\
  + \, t^{16} \, (2 + x^{ 4} + x^{18} + 2\,x^{22})  \\[1.3 mm]
& + \, t^{17} \, (1 - x^{12}) %\\
  + \, t^{18} \, (1 + x^{ 2} + x^{18} + x^{20}) %\\
  + \, t^{19} \, (1 - x^{28})  \\[1.3 mm]
& + \, t^{20} \, (2-x^{ 6}-x^{12}+2\,x^{18}-x^{24}-x^{30}) %\\[1.3 mm]
  + \, t^{21} \, (1 - x^{ 8}) \\[1.3 mm]
& + \, t^{22} \, (1 + x^{16} + x^{18} + x^{34}) %\\[1.3 mm]
  + \, t^{23} \, (1 - x^{24})  \\[1.3 mm]
& + \, t^{24} \, (2 + 2\,x^{14} + x^{18} + x^{32})  %\\[1.3 mm]
  + \, t^{25} \, (1 - x^{ 4})  %\\[1.3 mm]
  + \, t^{26} \, (1 - x^{ 6} - x^{24} + x^{30})  \\[1.3 mm]
& + \, t^{27} \, (1 - x^{20}) %\\[1.3 mm]
  + \, t^{28} \, (1 + x^{10} + x^{18} + x^{28})  \\[1.3 mm]
& + \, t^{30} \, (1 + x^{26}) %\\[1.3 mm]
  + \, t^{40} \, (1 + x^{34}) %\\[1.3 mm]
   \} \, (1-x^{36})^{-1}  \,.  \end{array}\ee

In the presence of fixed points the above prescription for
obtaining the \epop\ is not yet
quite complete, since from the quantum numbers of a fixed point
alone it cannot be decided whether a field into which the
fixed point is resolved and which appears in the orbit of
another \rgs\ is a \rgs\ (or the superpartner of a \rgs) or not.
In principle one could resolve this ambiguity
by using the full $S$-matrix
of the theory to calculate the fusion rules which, in turn,
determine the orbits of the spinor current. But again, there is a way
to avoid this involved calculation, which has the additional benefit
of showing that the results for the \epop\
do not depend as strongly on the details of the resolution as
one might imagine.
To this end we note that an important check of the spectra obtained via the
\epop\ is provided by the results of \cite{butu}, where
an independent way to calculate the net generation number
\ng\ by means of the ordinary \pop\ $P(t)$ was found. Namely,
  \be \ng \equiv \ngen - \nagen = \frac1\mo \sum_{r,s = 0}^{\mo/2-1}
  P({\rm e}^{2 \pi\ii\, {\rm d}(r,s)/\mo}) \,,  \labl{but}
where ${\rm d}(r,s)$ stands for the largest common divisor
of the integers $r$ and $s$.

Now since \erf{but} determines the
net generation number \ng\ from the \opop\ alone, \ng\
cannot depend on the resolution procedure \cite{sche3}.
To determine the correct extended \pop, we thus simply have to
start with the most general
ansatz compatible with the prescriptions given above and calculate,
for all possible values of the unknown parameters that arise from the
orbits containing resolved fixed points, the net generation number for
string vacua that involve the model under investigation as one factor
theory. If the net generation number generated this way does not fit
the value prescribed by (\ref{but}), we can exclude the corresponding
set of values for the unknown parameters. (To resolve all
ambiguities uniquely, it is sometimes necessary to take into
account that we can apply all formulas not only
to tensor products with $c=9$, but to tensor products with
$c= 3 + 6n$ for any positive integer $n$ as well.)

As an example, let us have a look at the theory
$(BB, 4, 2)$ which has $c=9$. The analysis of the \rgs s shows
that the coefficient of $t$ in the \epop\ is the polynomial
  \be    14 + a_1 x^2 - a_2 x^4               \ee
(multiplied with the irrelevant factor $(1 + x^{6})^{-1})$.
Here $a_1$ and $a_2$ are parameters arising from the fixed point
ambiguities just described; they must be
integers between 4 and 7. Now \erf{but} shows that $\ng = 0$, so
that \erf{ep} yields $ a_1 + a_2=14$, which in the given range
has the unique solution $a_1 = a_2 = 7$. Once the exact form
of the extended \pop\ is known, we can read off the number
of generations and antigenerations separately, namely
$\ngen=\nagen=14.$

In fact, in most cases one deals with a tensor product of
\nn \cts\ rather than with a single theory. The \opop\
$P_{\rm tot}(t)$ of a tensor product is just the product
of the \opop s $P_i(t)$ of the factor theories. Concerning the \epop, it
must be noticed that for a tensor product the considerations above
are to be applied to the
total spinor current $S_{\rm tot}=S^{(0)}S^{(1)}\ldots S^{(n)}$;
$S_{\rm tot}$ is by definition the product of the spinor currents
$S^{(i)}$, $i=1,2,...\,,n,$  (given by \erf S) of
the $n$ factor theories, and of the spinor $S^{(0)}$ of the
$D_5$ WZW theory at level one that describes either part the
gauge sector of the string theory or bosonized fermions and superconformal
ghosts. (The simple current $S_{\rm tot}$
is just the space-time supersymmetry generator, and hence  for
space-time supersymmetric theories it should
be part of the chiral algebra of the theory, i.e.\ the modular invariant
to be chosen for the tensor product is just the corresponding simple
current invariant. This corresponds to the generalized GSO projection, and
in fact it is possible to recover the usual projection condition, namely
odd integer superconformal charges, from the projection
to integer monodromy charges with respect to $S_{\rm tot}$.)

To obtain the extended \pop\ ${\cal P}_{\rm tot}(t, x)$
of the total theory from the extended
\pop s ${\cal P}_{i}(t, x_i)$ of all factor theories, one has to implement the
following prescription \cite{sche3} that is consistent with
the multiplication of \opop s. Multiply all
 ${\cal P}_{i}(t, x_i)$, and remove afterwards all terms in which the
powers of the variables $x_i$ do not coincide. Finally replace the product
$x_1\cdot x_2\cdot\,...\,\cdot x_n$ by $x$.
This procedure implements the fact  that some power $(S_{\rm tot})^m$
of the spinor current maps a \rgs\ on the (superpartner of)
another \rgs\ iff the components of $(S_{\rm tot})^m$ do so
in any factor of the theory.

We present the results of our calculations in tables \ref{c=6}
to \ref{c=9}.
In table \ref{c=6} we list all tensor products that can be written as
the tensor product of a $c=6$ and of a \cc3 theory and
in which at least one factor is
neither a \hsc\ nor the model $(CC, 2, 1)$ that will be dealt with separately.
The un-numbered lines contain the relevant non-hermitian symmetric
theories, while
the numbered lines provide the spectra for those $c=9$ theories
that are obtained by tensoring the \cc6 part
with the following $c=3$ models, respectively:
  \be \begin{tabular}{l} 1 -- 1 -- 1\,, \\[1 mm]
  1 -- 4 \mbox{~~~~or~~~~$ (A, 1, 2, 3)$}\,,   \\[1 mm]
  2 -- 2   \,,  \\[1 mm] $(CC, 2, 1)$ \end{tabular}\labl{c3}
(the theories 1 -- 4 and $(A,1,2,3)$ possess the same \epop\
and therefore yield the same spectrum).  Here and below, the symbol
\,`--'\, is used to indicate the tensor product, and a single
integer $k$ stands for the \nn minimal model at level $k$.

Next we display, in table \ref{c6c},
all tensor products that contain the model $(CC, 2, 1)$
which has $c=3$, but do not contain any other non-hermitian
symmetric \ct. We can tensor this model
twice and use the five $c=3$ models listed in \erf{c3}; as
the model $(CC,2,1)$ itself occurs in that list,
this includes tensoring three copies of the model.
We can also tensor it with 17 different combinations of minimal models
and 27 other combinations of \hsc s with \cc6.
  \futnote{The list in \cite{foiq}, containing 28 \hsc s with $c=6$,
is incomplete in several respects. First, rather than
$(D, 5, 2)$ -- 16, one must use  the combinations $(D, 5, 2)$ and
$(D, 5, 1)$ -- 16. Further, it was not realized that the \ct\ \pcs{2}{2}
(appearing in three of the 28 theories) coincides with the minimal model at
level 8. Finally, the theories $(B, 3, 6)$ and $(B, 6, 3)$ which in
\cite{foiq} were supposed to be identical, are in fact \cite{sche3}
distinct \cfts. Implementing these corrections, the number of the models gets
reduced by one, leading to the correct number of 27 models.}
 Altogether, this yields $15\times5+4+17+27=123$
models with \cc9 that involve non-\hsc s and contain a \cc3 part.

Finally, in table \ref{c=9} we
list all tensor products having $c=9$ in which at least one factor is
not a \hsc\ and which do not contain a tensor product with
$c=3$. We find 75 models of this type.
The number of theories that we count as different gets reduced
by various identifications, to be discussed below,
among the total of 198 theories. We have taken care of
these identifications, thereby reducing the number of entries in
the tables \ref{c=6} to \ref{c=9} to 112.

\clearpg  {\small
\begin{table}[t] \caption{\cc9\ tensor product theories that
contain a \cc6 part combined with
a non-hermitian symmetric factor (different from (CC,2,1)), and
the associated generation and anti-generation numbers}
\label{c=6} \begin{center} \begin{tabular}{|r|l|r|r|r|} \hline
&&&&\\[-3.4 mm]
\multicolumn{1}{|c|} {\#} & \multicolumn{1}{c|} {Model}  &
\multicolumn{1}{c|} \ngen & \multicolumn{1}{c|} \nagen   &
\multicolumn{1}{c|} \ng \dhlinv{3.4}
  &   $(BA, 3, 1)$                                         &&&   \\[.5 mm]
\nxe & \spc        -- 2  -- 1 -- 1 -- 1    &  21 &  21 &$^*$ 0   \\[.5 mm]
\nxt & \spc        -- 2  -- 1 -- 4         &  31 &   7 &     24  \\[.5 mm]
\nxt & \spc        -- 2  -- 2 -- 2         &  39 &   3 &     36  \\[.5 mm]
\nxt & \spc        -- 2  -- $(CC, 2, 1)$   &  31 &   7 &     24  \\[1.8 mm]
  &   $(BA, 4, 1)$ or $(C3, 1)$ or $(\Gs, 2)$              &&&   \\[.5 mm]
\nxt & \spc              -- 1 -- 1 -- 1    &  21 &  21 &$^*$ 0   \\[.5 mm]
\nxt & \spc              -- 1 -- 4         &  31 &   7 &     24  \\[.5 mm]
\nxt & \spc              -- 2 -- 2         &  31 &   7 &     24  \\[.5 mm]
\nxt & \spc              -- $(CC, 2, 1)$   &  31 &   7 &     24  \\[1.8 mm]
  &   $(BB, 3, 1)$ or $(CC, 2, 3)$ or $(CC, 4, 1)$         &&&   \\[.5 mm]
\nxt & \spc        -- 1  -- 1 -- 1 -- 1    &  51 &   3 &     48  \\[.5 mm]
\nxt & \spc        -- 1  -- 1 -- 4         &  51 &   3 &     48  \\[.5 mm]
\nxt & \spc        -- 1  -- 2 -- 2         &  21 &  21 &$^*$ 0   \\[.5 mm]
\nxt & \spc        -- 1  -- $(CC, 2, 1)$   &  21 &  21 &$^*$ 0   \\[1.8 mm]
  &   $(BB, 4, 1)$ or $(CC, 2, 5)$ or $(CC, 6, 1)$         &&&   \\[.5 mm]
\nxt & \spc              -- 1 -- 1 -- 1    &  21 &  21 &$^*$ 0   \\[.5 mm]
\nxt & \spc              -- 1 -- 4         &  23 &  23 &     0   \\[.5 mm]
\nxt & \spc              -- 2 -- 2         &  44 &   8 &     36  \\[.5 mm]
\nxt & \spc              -- $(CC, 2, 1)$   &  23 &  23 &     0   \\[1.8 mm]
  &   $(CC, 2, 2)$ or $(CC, 3, 1)$ or $(\Gs, 1)$           &&&   \\[.5 mm]
\nxt & \spc        -- 3  -- 1 -- 1 -- 1    &  21 &  21 &$^*$ 0   \\[.5 mm]
\nxt & \spc        -- 3  -- 1 -- 4         &  21 &  21 &$^*$ 0   \\[.5 mm]
\nxt & \spc        -- 3  -- 2 -- 2         &  21 &  21 &$^*$ 0   \\[.5 mm]
\nxt & \spc        -- 3  -- $(CC, 2, 1)$   &  21 &  21 &$^*$ 0   \\[1.8 mm]
  &   $(CC, 3, 2)$                                         &&&   \\[.5 mm]
\nxt & \spc              -- 1 -- 1 -- 1    &  21 &  21 &$^*$ 0   \\[.5 mm]
\nxt & \spc              -- 1 -- 4         &  41 &   5 &    36   \\[.5 mm]
\nxt & \spc              -- 2 -- 2         &  41 &   5 &    36   \\[.5 mm]
\nxt & \spc              -- $(CC, 2, 1)$   &  41 &   5 &    36   \\[1.8 mm]
  &  $(\Gl, 1)  $                                          &&&   \\[.5 mm]
\nxt & \spc        -- 1  -- 1 -- 1 -- 1    &  29 &   5 &     24  \\[.5 mm]
\nxt & \spc        -- 1  -- 1 -- 4         &  29 &   5 &     24  \\[.5 mm]
\nxt & \spc        -- 1  -- 2 -- 2         &  21 &  21 &$^*$ 0   \\[.5 mm]
\nxt & \spc        -- 1  -- $(CC, 2, 1)$   &  21 &  21 &$^*$ 0
\\[1 mm] \hline
\end{tabular} \end{center}\end{table}

\clearpg
\begin{table}[t] \caption{\cc6 tensor products, and the net generation
number \ng\ for the \cc9 models obtained by tensoring in addition
with $(CC,2,1)$}
\label{c6c} \begin{center} \begin{tabular}{|r|l|r|r|r|} \hline
&&&&\\[-3.4 mm]
\multicolumn{1}{|c|} {\#} & \multicolumn{1}{c|} {Model (\cc6 part)}  &
\multicolumn{1}{c|} \ngen & \multicolumn{1}{c|} \nagen   &
\multicolumn{1}{c|} \ng \dhlinv{3.4}
 \nxe &      1 --  1 -- 1 -- 1 -- 1 -- 1  &  21 &  21 &$^*$  0  \\[.35 mm]
 \nxt &      1 --  1 -- 1 -- 1 -- 4       &  35 &  11 &     24  \\[.35 mm]
 \nxt &      1 --  1 -- 1 -- 2 -- 2       &  21 &  21 &$^*$  0  \\[.35 mm]
 \nxt &      1 --  1 -- 2 -- 10           &  35 &  11 &     24  \\[.35 mm]
 \nxt &      1 --  1 -- 4 --  4           &  51 &   3 &     48  \\[.35 mm]
 \nxt &      1 --  2 -- 2 --  4           &  51 &   3 &     48  \\[.35 mm]
 \nxt &      2 --  2 -- 2 --  2           &  61 &   1 &     60  \\[.35 mm]
 \nxt &      1 --  5 -- 40                &  35 &  35 &      0  \\[.35 mm]
 \nxt &      1 --  6 -- 22                &  43 &  19 &     24  \\[.35 mm]
 \nxt &      1 --  7 -- 16                &  43 &  19 &     24  \\[.35 mm]
 \nxt &      1 --  8 -- 13                &  27 &  27 &      0  \\[.35 mm]
 \nxt &      1 --  10 -- 10               &  59 &  11 &     48  \\[.35 mm]
 \nxt &      2 --  3 -- 18                &  39 &  15 &     24  \\[.35 mm]
 \nxt &      2 --  4 -- 10                &  45 &   9 &     36  \\[.35 mm]
 \nxt &      2 --  6 --  6                &  55 &   7 &     48  \\[.35 mm]
 \nxt &      3 --  3 --  8                &  39 &  15 &     24  \\[.35 mm]
 \nxt &      4 --  4 --  4                &  60 &   6 &     54  \\[.35 mm]
 \nxt &      \pcs{2}{4} --  12            &  38 &  20 &     18  \\[.35 mm]
 \nxt &      \pcs{2}{5} --   6            &  55 &   7 &     48  \\[.35 mm]
 \nxt &      \pcs{2}{6} --   1 --  1      &  21 &  21 &$^*$  0  \\[.35 mm]
 \nxt &      \pcs{2}{6} --   4            &  23 &  23 &      0  \\[.35 mm]
 \nxt &      \pcs{2}{7} --   3            &  39 &  15 &     24  \\[.35 mm]
 \nxt &      \pcs{2}{9} --   2            &  45 &   9 &     36  \\[.35 mm]
 \nxt &      \pcs{2}{15} --  1            &  43 &  19 &     24  \\[.35 mm]
 \nxt &      \pcs{3}{3} --   5            &  21 &  21 &$^*$  0  \\[.35 mm]
 \nxt &      \pcs{3}{4} --   2            &  51 &   3 &     48  \\[.35 mm]
 \nxt &      \pcs{3}{5} --   1            &  21 &  21 &$^*$  0  \\[.35 mm]
 \nxt &      \pcs{3}{8}                   &  45 &   9 &     36  \\[.35 mm]
 \nxt &      \pcs{4}{5}                   &  41 &   5 &     36  \\[.35 mm]
 \nxt & $(A, 2, 2, 4)$                    &  51 &   3 &     48  \\[.35 mm]
 \nxt & $(B, 6, 3)$                       &  21 &  21 &$^*$  0  \\[.35 mm]
 \nxt & $(C, 2, 3)$        --   2         &  51 &   3 &     48  \\[.35 mm]
 \nxt & $(C, 2, 6)$                       &  23 &  23 &      0  \\[.35 mm]
 \nxt & $(C, 3, 2)$                       &  21 &  21 &$^*$  0  \\[.35 mm]
 \nxt & $(C, 4, 1)$        --   1         &  35 &  11 &     24  \\[.35 mm]
 \nxt & $(D, 5, 2)$                       &  21 &  21 &$^*$  0  \\[.35 mm]
 \nxt & $(CC, 2, 1)$ -- 1 -- 1 -- 1       &  21 &  21 &$^*$  0  \\[.35 mm]
 \nxt & $(CC, 2, 1)$ -- 1 -- 4            &  51 &   3 &     48  \\[.35 mm]
 \nxt & $(CC, 2, 1)$ -- 2 -- 2            &  51 &   3 &     48  \\[.35 mm]
 \nxt & $(CC, 2, 1)$ -- $(CC, 2, 1)$      &  51 &   3 &     48
 \\[1 mm] \hline
\end{tabular} \end{center}\end{table}

\clearpg \begin{table}[t]
\caption{\cc9 tensor products that contain a non-\hsc\ and cannot
be decomposed in the tensor product of a \cc3 and a \cc6 theory}
\label{c=9} \begin{center} \begin{tabular}{|r|l|r|r|r|} \hline
&&&&\\[-3.4 mm]
\multicolumn{1}{|c|} {\#} & \multicolumn{1}{c|} {Model}  &
\multicolumn{1}{c|} \ngen & \multicolumn{1}{c|} \nagen   &
\multicolumn{1}{c|} \ng \dhlinv{3.4}
    & $(BA, 3,  1)$                                       &&&  \\[.5 mm]
\nxe& \spd          --  1 -- 1  -- 10     &  19 &  19 &     0  \\[.5 mm]
\nxt& \spd          --  3 -- 18           &  23 &  23 &     0  \\[.5 mm]
\nxt& \spd          --  4 -- 10           &  27 &  15 &    12  \\[.5 mm]
\nxt& \spd          --  6 --  6           &  35 &  11 &    24  \\[.5 mm]
\nxt& \spd          -- $(A, 1, 2, 9)$     &  27 &  15 &    12  \\[.5 mm]
\nxt& \spd          -- $(A, 1, 3, 4)$     &  31 &   7 &    24  \\[.5 mm]
\nxt& \spd          -- $(B, 6, 2)$        &  35 &  11 &    24  \\[.5 mm]
\nxt& \spd          -- $(C, 2, 3)$        &  31 &   7 &    24  \\[.5 mm]
\nxt& \spd          -- $(BA, 3, 1)$       &  15 &  15 &     0  \\[.5 mm]
\nxt& $(BA, 3,  2)$ -- 12                 &  12 &   8 &     4  \\[.5 mm]
\nxt& $(BA, 3,  4)$                       &  14 &   2 &    12  \\[.5 mm]
\nxt& $(BA, 5,  1)$ --  2                 &  15 &  15 &     0  \\[.5 mm]
\nxt& $(BA, 6,  1)$                       &  15 &  15 &     0  \\[.5 mm]
    & $(BB, 3,  1)$ or $(CC, 2, 3)$ or $(CC, 4, 1)$       &&&  \\[.5 mm]
\nxt& \spd          --  2 -- 10           &  35 &  11 &    24  \\[.5 mm]
\nxt& \spd          --  4 --  4           &  43 &   7 &    36  \\[.5 mm]
\nxt& \spd          -- $(A, 1, 2, 6)$     &  35 &  11 &    24  \\[.5 mm]
\nxt& \spd          -- $(A, 2, 2, 2)$     &  51 &   3 &    48  \\[.5 mm]
\nxt& $(BB, 3,  3)$                       &  17 &   5 &    12  \\[.5 mm]
\nxt& $(BB, 4,  2)$                       &  14 &  14 &    0   \\[.5 mm]
    & $(BB, 5,  1)$ or $(CC, 2, 7)$ or $(CC, 8, 1)$        &&& \\[.5 mm]
\nxt& \spd         --  8                  &  39 &  15 &   24  \\[.5 mm]
    & $(BB, 6,  1)$ or $(CC, 2, 9)$ or $(CC, 10, 1)$      &&& \\[.5 mm]
\nxt& \spd          -- 1  -- 1            &  29 &  29 &    0  \\[.5 mm]
\nxt& \spd          -- 4                  &  44 &  14 &   30  \\[.5 mm]
    & $(BB, 8,  1)$ or $(CC, 2, 13)$ or $(CC, 14, 1)$   &&& \\[.5 mm]
\nxt& \spd          -- 2                  &  34 &  34 &    0  \\[.5 mm]
    & $(BB, 12, 1)$ or $(CC, 2, 21)$ or $(CC, 22, 1)$   &&& \\[.5 mm]
\nxt& \spd          -- 1                  &  43 &  43 &    0  \\[.5 mm]
    & $(CC, 2, 2)$ or $(CC, 3, 1)$ or $(\Gs, 1)$          &&&  \\[.5 mm]
\nxt& \spd         -- 1 -- 1 -- 28        &  23 &  23 &     0  \\[.5 mm]
\nxt& \spd         -- 4 -- 28             &  29 &  29 &     0  \\[.5 mm]
\nxt& \spd         -- 8 -- 8              &  47 &  11 &    36  \\[.5 mm]
\nxt& \spd         -- $(A, 1, 2, 12)$     &  35 &  17 &    18  \\[.5 mm]
\nxt& \spd         -- $(B, 8, 2)$         &  41 &   5 &    36  \\[.5 mm]
    & $(CC, 2, 4)$ or $(CC, 5, 1)$                        &&& \\[.5 mm]
\nxt& \spd         -- $(A, 1, 2, 4)$      &  29 &  14 &   15
 \\[1 mm] \hline
\end{tabular} \end{center}\end{table}
  \clearpg

\addtocounter{table}{-1}\begin{table}[t] \caption{{\em continued.}}
\begin{center} \begin{tabular}{|r|l|r|r|r|} \hline
&&&&\\[-3.4 mm]
\multicolumn{1}{|c|} {\#} & \multicolumn{1}{c|} {Model}  &
\multicolumn{1}{c|} \ngen & \multicolumn{1}{c|} \nagen   &
\multicolumn{1}{c|} \ng \dhlinv{3.4}
    & $(CC, 2, 6)$ or $(CC, 7, 1)$                        &&& \\[.5 mm]
\nxt& \spd         -- 16                  &  27 &  27 &    0  \\[.5 mm]
\nxt& $(CC, 3, 5)$ or $(CC, 6, 2)$        &  20 &  20 &    0  \\[.5 mm]
\nxt& $(CC, 4, 3)$                        &  29 &   9 &   20  \\[.5 mm]
\nxt& $(C4, 1)$                           &  15 &   7 &    8  \\[.5 mm]
\nxt& $(D4, 1)$     --  $(A, 1, 2, 4)$    &  23 &  11 &   12  \\[.5 mm]
\nxt& $( \dfe , 1)$                       &  12 &   0 &   12  \\[.5 mm]
\nxt& $( \dfz , 1)$ --  16                &  19 &  19 &    0  \\[.5 mm]
\nxt& $(F4, 1)$     --   8                &  25 &  13 &   12  \\[.5 mm]
    & $(\Gl, 1)$                                         &&&  \\[.5 mm]
\nxt& \spd            -- 2 -- 10          &  17 &  17 &    0  \\[.5 mm]
\nxt& \spd            -- 4 --  4          &  23 &  11 &   12  \\[.5 mm]
\nxt& \spd            -- $(A, 1, 2, 6)$   &  17 &  17 &    0  \\[.5 mm]
\nxt& \spd            -- $(A, 2, 2, 2)$   &  29 &   5 &   24  \\[.5 mm]
\nxt& $(\Gl, 2)$      -- 7                &   9 &   9 &    0  \\[.5 mm]
\nxt& $(\Gs, 5)$                          &  20 &  20 &    0
 \\[1 mm] \hline \end{tabular} \end{center}\end{table}
 }

\subsection{Level-rank duality}
As it turns out, the extended \pop s for several theories that
are defined as distinct naive \cts\ coincide.
The cases where this happens can be easily read off the tables
as follows. If the \epop s of some theories are identical,
these theories are listed together in an un-numbered line;
the numbered line(s) following this line then contain the
theories with which each of them can be tensored to obtain a \cc9 theory.
For instance, the line preceding the lines numbered from 25 to 29
in table \ref{c=9} shows that the theories $(CC,2,2)$, $(CC,3,1)$ and
$(\Gs,1)$ have identical \epop s.
We have also taken into account the known \cite{sche3} fact that the
\epop s of the \hsc s $(A,1,2,3)$, $(A,2,2,2)$,
 \futnote{However, in the table \ref{c=9} we have nevertheless
kept the entries \# 17 and \# 42 containing $(A,2,2,2)$, because after
identification with \,1\,--\,1\,--\,1\,--\,1\,, they would
correspond to entries in a different table, namely table \ref{c=6}.}
 $(C,3,1)$, and $(D,5,1)$
coincide with those of the tensor products
\,1\,--\,4\,, \,1\,--\,1\,--\,1\,--\,1\,, \,3\,--\,3\,,
and \,1\,--\,7\, of minimal models, respectively.

{}From the experience with coset constructions,
the observation that there exist a priori distinct
\cts\ with coinciding \epop s is not very spectacular.
What is surprising, however, is that in fact for {\em all\/}
non-hermitian \nn \cts\ for which the \opop s
are identical (compare table \ref{pop} above),
the same is true for the \epop s.

In particular, the \epop s of the two theories $(CC, r, k)$\, and\,
$(CC, k+1, r -1)$ for $r \geq 2$ are identical for all values
of $r$ and $k$ for which we calculated them. Since the
extended \pop\ describes explicitly
also part of the structure of the chiral ring
(whereas the ordinary \pop\ essentially counts multiplicities),
this is in itsef already a rather strong hint that these theories
should be closely related, if not be identical as \cfts.

When looking at the \cft\ beyond the chiral ring one realizes that the
number of primary fields in both theories is identical.
But we can do even better and construct a map between
primary fields of these theories which can be shown to preserve
most, if not all, of the \cft\ properties.
Namely, the $\hat h'=(C_k)_r$-weight $\lambda'$ of a
(representative of a) primary field
of $(CC, k+1, r -1)$ is related to the $g=(C_r)_k$-weight $\Lambda$
of the associated field of $(CC, r, k)$ by
  \be  \Lambda\mapsto\lambda':\qquad Y^{}_{\lambda'}=
  (Y^{}_{\Lambda})^{\rm c}_{} ,  \ee
where $Y_\Lambda$ denotes the Young tableau associated to the
highest  weight $\Lambda$, and the symbol `\,$^{\rm c}$\,' stands for
the operation of first taking the complement of $Y_\Lambda$ with respect
to the rectangular Young tableau $Y^{}_{k\Lambda_{(r)}}$ and then
reflecting the tableau so obtained along an axis perpendicular to the
main diagonal in such a way as to obtain a standard Young tableau
of $(C_k)_r$. This map is well-known \cite{mnrs} from the so-called
`level-rank duality' of the $(C_r)_k$
and $(C_k)_r$ \wzwts. To obtain the duality map for the \cts,
it has to be supplemented by an analogous relation between
the $g'$- and $\hat h$-weights, and by prescriptions for the \sod\ and
\ue\ part which include in particular
  \be  Q'= \left\{ \begin{array}{lll}
  -Q & {\rm for} & {\rm x}\in\{{\rm s,c}\}, \\[1 mm]
  k+r+1-Q & {\rm for} & {\rm x}\in\{{\rm v,0}\}.
  \end{array}\right.  \ee

In fact we expect that the \cts\ $(CC, r, k)$ and $(CC, k+1, r -1)$
are merely two different descriptions of one and
the same \cft, so that there is even no need
for a marginal flow to interpolate between them. We will come back
to this issue in a forthcoming note.

\sect{Conclusions}
In this paper we have presented a detailed analysis of non-hermitian
symmetric \nn superconformal coset theories and of
compactifications of the heterotic string that contain such
\cts\ in their inner sector. In addition, we have
proven some general statements on the structure of any \nn \ct.
Concerning
the non-hermitian symmetric \cts\ themselves, we have shown that they indeed
allow for an
interpretation as a consistent \cft; this lends further
support to the expectation that {\em any\/} \ct, naively
`defined' as \cosetk gh, possesses such an interpretation.
In particular, it was shown that the fixed points that arise in
the process of field identification can be resolved by
the methods of \cite{scya5}.

The spectra of string compactifications that we obtained
are certainly not spectacular, but rather
similar to those obtained for previously analyzed
classes of compactifications. This confirms the
by now common lore that extending the set of
string compactifications does not have a very large impact
on the set of known spectra.
The results also confirm the experience that when employing
more complicated \cfts, the numbers of generations and
anti-generations tend to be smaller than in the case of
simpler (say, \nn minimal) theories.

There still remain several directions for further work on the
subject. First, one may consider modular invariant
combinations of characters of the $g$- and $h$-\,\wzwts\ other
than the diagonal one,
in particular non-diagonal invariants of tensor product
theories that are not obtained from products of the invariants of the
affine Lie algebras associated to the individual factor
theories. One may also investigate whether the \cts, or at least their tensor
products with $c=3n$, might have a description in
terms of \lagi\ potentials or Calabi\hy Yau manifolds, or of orbifolds
thereof (while it is generally assumed that such a connection
should exist, the arguments supporting this expectation are
far from being rigorous). To identify these different descriptions
it would be very
useful to have a more detailed knowledge of the discrete symmetries of
the models. One of these discrete symmetries is obvious,
namely the symmetry of the operator products
induced by conservation of the superconformal \ue-charge;
but generically there may be further symmetries, and it is not
clear how one could find all of them.
Of course, once discrete symmetries are known,
one can divide out some of them  so as to obtain orbifolds of our models.

We also mention that a complete computation of massless string spectra,
i.e.\ including the
fields that are singlets under $E_6$, would clearly be welcome.
To this end one would have to compute
the character decompositions by means of the Kac\hy Weyl character formula
(in order to identify null states and to obtain the integer part
of the conformal weight of a field), and implement the beta vector method
known from tensor products of minimal models.
It is evident that this is a laborious
procedure, and any alternative method would be of great interest.

Another interesting aspect of the string spectra obtained in the
paper is that the \epop s \ptx, and hence the generation numbers
\ngen\ and \nagen\ of the associated string compactifications,
of two theories are identical whenever the \opop s $P(t)={\cal P}
(t,0)$ are. This indicates that the structure of the \epop\ is
to a large extent already dictated by the
information contained in the {\em ordinary\/} \pop; in particular
(compare \cite{sche3}), in the presence of fixed points
the numbers of massless generations and anti-generations do not
depend at all on the details of the resolution procedure.
A general proof of this observation is however still lacking.

Let us finally come back to the hypothesis that, given a chain of subalgebras
$ h_1 \hookrightarrow h_2 \hookrightarrow g $,
the \ct\ \cosetk{g}{h_1} should correspond to the tensor product of
the two cosets \cosetk{g}{h_2} and \coset{h_2}{h_1}{k'}, with a suitably
chosen non-product modular invariant. We emphasize that
in the presence of fixed points this hypothesis is far from being proven.
With the methods employed in the present paper it should be
straightforward to examine the structure of both \cosetk{g}{h_1}
and the tensor product of \cosetk{g}{h_2} and \coset{h_2}{h_1}{k'}
in detail, and thereby test the hypothesis for any given chain of
embeddings. To prove the equivalence in full generality,
however, still a deeper understanding of the structure of coset
\cfts\ seems to be necessary.

\clearpg\appendix \sect{Appendix: Hasse diagramms}
The Hasse diagram \cite{BAea} for an embedding \hing\ of a reductive
Lie algebra in a simple Lie algebra
is the graph of the coset $\wg/\wh$, interpreted as
a subgraph of the graph of \wg, with the edges as prescribed
by the Bruhat ordering of \wg.
(Hasse diagrams also arise in the description of the topological
structure of generalized flag manifolds and of the structure of the
\bgg-resolution of Verma modules.) The nodes of the Hasse diagram
correspond to those representatives of elements of \wgh\ that
send a dominant
$g$-weight $\Lambda$ to a dominant $\hat h$-weight $\lambda$,
i.e.\ to $h$-positive elements of \wg, and the integer $i$
attached to an edge indicates that the two nodes connected
by the edge correspond to Weyl group elements $w$ and $w'$
related by $w'=w_{(i)}w$, with $w_{(i)}$ the $i$th
fundamental reflection.
For an embedding \hing\ for which the Dynkin diagram of $h$
is obtained by deleting the node with label $\xx$ from the Dynkin
diagram of $g$,
the Hasse diagram is isomorphic to the \wg-orbit of \lx,
i.e.\ to the `restricted weight diagram' that one obtains when
acting successively on the weight \lx\ with the fundamental reflections.

The Hasse diagrams for the embeddings relevant to \hsc s have been
described in \cite{ekmy}. Below we present the Hasse diagrams
for some of the non-hermitian symmetric cases which appear in
table \ref{series} (the diagrams for the remaining cases, i.e.\ the
$BA$ and $BB$ series and the two $D5$ theories look more
complicated, and we refrain from drawing them here).
 \futnote{The Hasse diagram of \Wgh{F_4}{C_3}\
can also be found in \cite[p.\,86]{coll}.}

\vskip 3 mm Hasse diagram of \Wgh{C_n}{C_{n-1}}: \vskip 15 mm

\begin{picture}(260,60)(0,-80)
\mpcis00{30}045 \mpcis{120}04061 \mpcis{170}0{30}045
\mpcis{290}04061 \mpcis{340}0{30}035 \mplin00{150}0210{110}
\pulin{260}010{20} \pulin{320}010{80}
\mpsss{12}2{370}021 \mpsss{42}2{310}022 \putss{72}23
\mpsss{179}2{60}02{$n$--1} \putss{214}2{$n$}
\end{picture}

\vskip-18 mm Hasse diagram of \Wgh{C_3}{A_1\op A_1}: \vskip 5.6 mm

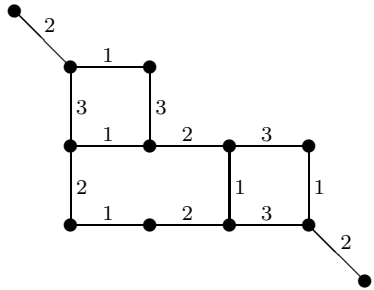
\begin{picture}(260,160)(-80,-80)
\mpcis{30}0{30}045 \mpcis{30}{30}{30}045 \mpcis{30}{60}{30}025
\mpcis{8.8}{81.2}{132.4}{-102.4}25
\mplin{30}{30}{30}0201{30} \pulin{30}001{30} \mplin{90}0{30}0201{30}
\mplin{30}00{30}210{90} \pulin{30}{60}10{30}
\mplin{30}{60}{111.2}{-81.2}2{-1}1{21.2}
\mpsss{20}{73}{112}{-82}22 \mpsss{42}20{30}31 \mpsss{72}20{30}22
\mpsss{102}20{30}23 \putss{32}{12}2 \mpsss{32}{42}{30}023
\mpsss{92}{12}{30}021
\end{picture}

%newpage
\vskip -11 mm Hasse diagram of \Wgh{C_4}{A_2\op A_1}: \vskip .2 mm

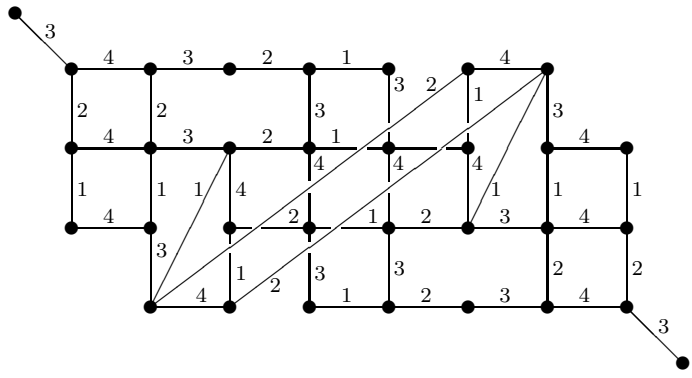
\begin{picture}(270,110)(-50,20)
\mpcis{8.8}{111.2}{252.4}{-132.4}25 \mpcis{30}{90}{30}075
\mpcis{30}{60}{30}085 \mpcis{30}{30}{30}085 \mpcis{60}0{30}075
\mplin{30}{90}{231.2}{-111.2}2{-1}1{21.2}
\mplin{30}{90}{90}{-90}210{120}  \mplin{30}{60}{102}{-30}210{108}
\mplin{142}{60}{-40}{-30}210{26} \mplin{172}{60}{-84}{-30}210{10}
\mplin{30}{30}{30}{-30}210{30}   \mplin{180}{90}{30}{-30}210{30}
\pulin{30}{30}01{60} \pulin{240}001{60} \mplin{60}0{150}0201{90}
\mplin{60}0{120}{30}212{30} \mplin{60}0{30}0243{120}
\mpsss{20}{102}{232}{-112}23 \mpsss{42}{32}0{30}34
\mpsss{72}{62}0{30}23 \mpsss{102}{62}0{30}22 \mpsss{132}20{90}21
\mpsss{162}20{30}22   \mpsss{192}20{30}23    \mpsss{222}20{30}34
\mpsss{32}{42}{30}021 \mpsss{32}{72}{30}022  \mpsss{122}{72}{90}023
\putss{92}{42}4       \mpsss{212}{42}{30}021 \mpsss{212}{12}{30}022
\mpsss{121.5}{52}{30}034\putss{77}24 \putss{164}{82}2
\putss{105}{5.8}2 \putss{76}{41.8}1  \putss{92}{10}1
\putss{112}{32}2  \putss{142}{32}1   \putss{128}{62}1
\putss{192}{92}4  \putss{122}{10}3   \putss{152}{12}3
\putss{152}{82}3  \putss{188.5}{42}1 \putss{62}{19}3
\putss{182}{78}1
\pulin{90}001{20}        \pulin{90}{60}0{-1}{36}  \pulin{120}001{20}
\pulin{120}{25}01{18}    \pulin{120}{90}0{-1}{42} \pulin{150}001{43}
\pulin{150}{47.5}01{18.5}\pulin{150}{90}0{-1}{20}
\pulin{180}{30}01{35.5}  \pulin{180}{90}0{-1}{20}
\end{picture}

\clearpg

\vskip -7 mm Hasse diagram of \Wgh{D_4}{A_1\op A_1\op A_1}: \vskip 18 mm

\begin{picture}(260,220)(-80,-80)
\mpcis{45}{126.2}{30}035   \mpcis{30}{141.2}{30}025
\mpcis{30}{111.2}{30}025   \mpcis{45}{66.2}{15}{15}35
\mpcis{66.2}{45}{15}{15}35 \mpcis{45}{96.2}{51.2}{-51.2}25
\mpcis{126.2}{45}0{30}35   \mpcis{112.2}{30}0{30}25
\mpcis{141.2}{30}0{30}25   \mpcis{8.8}{162.4}{153.6}{-153.6}25
\mplin{8.8}{162.4}{117.4}{-117.4}21{-1}{36.2}
\mplin{45}{126.2}{36.2}{-66.2}210{60}    \pulin{66.2}{45}10{30}
\mplin{30}{141.2}{15}{-45}210{30} \mplin{96.2}{75}{15}{-45}210{30}
\mplin{45}{66.2}{81.2}{8.8}201{30}\mplin{30}{111.2}{15}{-15}201{30}
\mplin{60}{141.2}{36.2}{-96.2}21{-1}{15} \pulin{75}{96.2}01{30}
\mplin{111.2}{30}{30}0201{30}     \mplin{30}{111.2}{30}021{-1}{15}
\mplin{96.2}{75}{30}021{-1}{15}   \mplin{45}{66.2}{15}{15}21{-1}{21.2}
\mplin{60}{111.2}0{17}201{13}     \mplin{75}{96.2}{30}{30}21{-1}{21.2}
\mplin{60}{81.2}0{17}201{13}      \mplin{96.2}{45}0{17}201{13}
\mplin{126.2}{45}0{17}201{13}     \mplin{30}{111.2}{17}0210{13}
\mplin{96.2}{45}{17}0210{13}
\putss{42}{143}1 \putss{36}{112.7}1 \putss{143}{42}1
\putss{51.4}{50.4}1 \putss{112.6}{35}1 \mpsss{52}{97.7}0{30}21
\mpsss{97.6}{50}{30}021 \putss{87}{128}2 \putss{79.5}{38.9}2
\putss{39.1}{78.5}2 \putss{128}{87}2 \putss{61.8}{87}2
\putss{86.4}{85.6}2 \putss{86.6}{61.6}2
\mpsss{20}{152.6}{132.4}{-132.4}22 \putss{24.6}{123.4}3
\putss{77}{108.4}3 \putss{124.5}{24.1}3 \putss{116.4}{115.6}3
\putss{108.6}{76.4}3 \mpsss{46.4}{117}{15}023
\mpsss{117}{46.4}0{15}23 \putss{67.1}{104.1}4 \putss{33}{98.7}4
\putss{100}{32}4 \putss{134}{37.2}4 \putss{71.4}{70.4}4
\mpsss{38.3}{132.9}{30}024 \mpsss{104.6}{67}{30}024
\end{picture}

%newpage
Hasse diagram of \Wgh{F_4}{C_3}: \vskip 8 mm

\begin{picture}(260,320)(-30,-80)
\mpcis{90}{180}{30}035 \mpcis{90}{150}{30}035 \mpcis{120}{120}{30}055
\mpcis{120}{90}{30}055 \mpcis{210}{60}{30}025
\mpcis{26.4}{243.6}{21.2}{-21.2}35 \mpcis{261.2}{38.8}{21.2}{-21.2}35
\mplin{120}{120}0{-30}210{120} \mplin{90}{180}0{-30}210{60}
\pulin{210}{60}10{30} \mplin{120}{120}{30}0201{60}
\pulin{90}{150}01{30} \mplin{120}{90}{60}0201{30}
\mplin{210}{60}{30}0201{60} \mplin{90}{180}{213.6}{-183.6}2{-1}1{63.6}
\mpsss{38}{235}{256}{-226}21 \mpsss{59}{214}{214}{-184}22
\mpsss{80}{193}{172}{-142}23 \mpsss{92}{162}{30}034
\mpsss{122}{132}{30}023 \putss{122}{102}2  \mpsss{182}{102}{30}031
\mpsss{212}{72}{30}022  \mpsss{102}{152}0{30}22 \mpsss{132}{92}0{30}41
\mpsss{162}{92}0{30}22  \mpsss{192}{92}0{30}23  \mpsss{222}{62}0{30}34
\end{picture}

\clearpg

\vskip -17 mm Hasse diagram of \Wgh{G_2}{A_1^>}: \vskip 16 mm

\begin{picture}(160,60)(-120,-80)
\mpcis00{30}065 \pulin0010{150} \mpsss{12}2{60}032 \mpsss{42}2{60}021
\end{picture}

\vskip-9 mm Hasse diagram of \Wgh{G_2}{A_1^<}: \vskip 16 mm

\begin{picture}(160,60)(-120,-80)
\mpcis00{30}065 \pulin0010{150} \mpsss{12}2{60}031 \mpsss{42}2{60}022
\end{picture}

{\bf Acknowledgement.} It is a pleasure to thank M.\ Kreuzer,
A.N.\ Schellekens and M.G.\ Schmidt for interesting discussions, and
the theory group of NIKHEF for hospitality.
Some of the manipulations of extended \pop s were
performed with the symbolic program FORM \cite{FORM}. \\[5 mm]

\mbox{}

 \end{document}